\documentclass[5p,times]{elsarticle}
\usepackage{hyperref}

\journal{Information Sciences}

\usepackage{balance}
\usepackage{listings}
\usepackage{color}

\definecolor{dkgreen}{rgb}{0,0.6,0}
\definecolor{gray}{rgb}{0.5,0.5,0.5}
\definecolor{mauve}{rgb}{0.58,0,0.82}

\makeatletter
\def\lst@makecaption{%
  \def\@captype{table}%
  \@makecaption
}
\makeatother

\graphicspath{{img/}}

\usepackage{url}
\usepackage{multirow}

\usepackage{amssymb}
\usepackage{amsmath}
\newtheorem{definition}{Definition}
\newtheorem{example}{Example}  %gws
\newtheorem{strategy}{Strategy}    % gws
\newtheorem{proof}{Proof}    % gws
\newtheorem{lemma}{Lemma}  %gws
\usepackage{algorithm}
\usepackage{algorithmic}
\usepackage{array}

\usepackage[caption=false,font=footnotesize,labelfont=sf,textfont=sf]{subfig}

\begin{document}

\begin{frontmatter}

\title{Discovering High Utility-Occupancy Patterns from Uncertain Data}

\author{Chien-Ming Chen$ ^{1} $,
	Lili Chen$ ^{1} $,
	Wensheng Gan$ ^{2}$*,
	Lina Qiu$ ^{3} $,
	Weiping Ding$ ^{4} $
}

\address{$ ^{1} $College of Computer Science and Engineering, Shandong University of Science and Technology, Qingdao 266590, Shandong, China} 
\address{$ ^{2} $College of Cyber Security, Jinan University, Guangzhou 510632, Guangdong, China}
\address{$ ^{3} $School of Software, South China Normal University, Foshan 528200, Guangdong, China}

\address{$ ^{4} $School of Information Science and Technology, Nantong University, Nantong 226019, Jiangsu, China}

\address{Email: chienmingchen@ieee.org, lilichien3@gmail.com,  wsgan001@gmail.com, lina.qiu@scnu.edu.cn, dwp9988@163.com} % psyu@uic.edu}

\cortext[cor1]{Corresponding author. Email: wsgan001@gmail.com} % Tel.: +86-XXXX.}

\begin{abstract}

It is widely known that there is a lot of useful information hidden in big data, leading to a new saying that "data is money." Thus, it is prevalent for individuals to mine crucial information for utilization in many real-world applications. In the past, studies have considered frequency. Unfortunately, doing so neglects other aspects, such as utility, interest, or risk. Thus, it is sensible to discover high-utility itemsets (HUIs) in transaction databases while utilizing not only the quantity but also the predefined utility. To find patterns that can represent the supporting transaction, a recent study was conducted to mine high utility-occupancy patterns whose contribution to the utility of the entire transaction is greater than a certain value. Moreover, in realistic applications, patterns may not exist in transactions but be connected to an existence probability. In this paper, a novel algorithm, called \underline{H}igh-\underline{U}tility-\underline{O}ccupancy \underline{P}attern \underline{M}ining in \underline{U}ncertain databases (UHUOPM), is proposed. The patterns found by the algorithm are called \underline{P}otential \underline{H}igh \underline{U}tility \underline{O}ccupancy \underline{P}atterns (PHUOPs). This algorithm divides user preferences into three factors, including support, probability, and utility occupancy. To reduce memory cost and time consumption and to prune the search space in the algorithm as mentioned above, probability-utility-occupancy list (PUO-list) and probability-frequency-utility table (PFU-table) are used, which assist in providing the downward closure property. Furthermore, an original tree structure, called support count tree (SC-tree), is constructed as the search space of the algorithm. Finally, substantial experiments were conducted to evaluate the performance of proposed UHUOPM algorithm on both real-life and synthetic datasets, particularly in terms of effectiveness and efficiency.

\end{abstract}

\begin{keyword}
  utility mining, utility occupancy, uncertain data, probability, potential pattern. % potential high utility occupancy pattern.
\end{keyword}

\end{frontmatter}

% \linenumbers

\section{Introduction}

With the prevalence of Internet of Things (IoTs) technology, information sensing equipment (such as sensors, RFID tags, and so on) always generate massive amounts of data per second. It is essential for humans to discover hidden and useful information from this rich data. Agrawal \textit{et al.} \cite{agrawal1994fast} first advocated the pioneering  Apriori algorithm to level-wisely discover frequent patterns from a precise database. Unfortunately, this method traverses the database multiple times and generates a host of candidate itemsets, which leads to too much memory and consumption during execution. Han \textit{et al.} \cite{han2004mining} further presented the FP-growth algorithm and invented a novel tree structure, named FP-tree, with which no candidates are generated. 

In recent decades, a multitude of investigators have hastened to improve data mining algorithms due to only finding limited interestingness measures, for example, frequency or support. However, these are insufficient because every object or item is actually unequal in the final analysis. Other preferences, like the profit, cost, risk, and weight \cite{gan2018survey,lin2016weighted}, have increasingly been studied, and they allow more valuable information to be discovered than the previous support-based mining algorithms. The utility-driven mining framework such as high-utility itemset mining (HUIM) model \cite{chan2003mining,yao2006mining} is thus proposed. HUIM considers the unit utility (external utility) and quantity (internal utility) of objects or items. Based on these two factors, it is easy to calculate the utility of itemsets. If the derived utility value is greater than a defined minimum utility threshold in advance, then the itemsets can be called high-utility itemsets (HUIs). After that, Liu \textit{et al.} \cite{liu2005two} designed the two-phase model in which the main emphasis is to find the upper bound of the utility of itemsets and then trim most of the unqualified itemsets without further calculating their supersets. Other utility-based mining algorithms, such as HUP-growth \cite{lin2011effective}, UP-growth \cite{tseng2010up}, HUI-Miner \cite{liu2012mining}, CoUPM \cite{gan2018coupm}, ProUM \cite{gan2019proum}, and HUSP-ULL \cite{gan2019fast} have also been proposed to deal with different mining tasks. Up to now, some studies of privacy-preserving utility mining also have been addressed and reviewed in prior work \cite{gan2018privacy}.

First, Tang \textit{et al.} \cite{tang2012incorporating} creatively proposed the concept of occupancy, which is the ratio of the number of items in the itemset to the one in the transaction, and considered it a dominant factor. Unfortunately, it could not help solve the problem of profit and utility. To address this issue, Shen \textit{et al.} \cite{shen2016ocean} blended the concepts of occupancy and utility, and then presented utility occupancy with the OCEAN algorithm, which is used to find patterns on behalf of the relative supporting transactions in the utility ratio. However, this utility-driven algorithm does not completely find the patterns that meet the requirements. Gan \textit{et al.} \cite{gan2019huopm} put forward a novel algorithm, called high utility-occupancy pattern mining (HUOPM) algorithm. It can avoid the errors in the OCEAN algorithm \cite{shen2016ocean} and uses two novel structures to prune the search space. Finding qualified utility occupancy patterns has a wide range of applications in real life, especially in the era of rapid information development. For instance, Alipay analyzes the consumption records of consumers on various online platforms, such as Taobao and Meituan, and calculates the proportion of goods consumed in the record to obtain products that can represent the corresponding trade and thus gain consumer spending habits to recommend products to consumers based on their preferences.

The algorithms mentioned above are all based on precise databases. Conversely, as a result of noisy data sources or failure of data transmission, losing some data is unavoidable and triggers uncertainty in databases \cite{aggarwal2009frequent}. Therefore, it is difficult to apply these existing algorithms to handle uncertain database. Uapriori \cite{chui2007mining} was the first algorithm to discover frequent itemsets from uncertain databases. It mainly adopts a generate-and-test mechanism. However, it is relatively unsatisfactory because of memory space consumption. Subsequently, an approach without generating candidates by utilizing a UFP-tree structure was introduced \cite{leung2008tree}. Utility-based mining in uncertain databases is also indispensable. For example, the upper-bound-based PHUI-UP \cite{lin2016efficient} algorithm and PU-list-based PHUI-List algorithm \cite{lin2016efficient} were designed to seek out potential high-utility patterns in uncertain databases. PHUI-UP adopts a hierarchical strategy and it depends on multiple scanning the database and generates a large number of candidate itemsets during the mining process. PHUI-List uses a vertical structure to store data, and the pruning strategy mentioned in this algorithm can speed up the mining process.

So far, decision-makers can not find out existing algorithms for analyzing some complicated data with uncertainty and revealing high utility-occupancy patterns. To address this issue, in this paper we focus on the need for combining pattern mining, utility occupancy, and uncertain databases. Thus we introduce a novel algorithm, called \underline{H}igh \underline{U}tility-\underline{O}ccupancy \underline{P}attern \underline{M}ining in \underline{U}ncertain databases (UHUOPM for short). Three factors are involved in this algorithm, namely, frequency, probability, and utility occupancy. 
Among them, frequency is mainly applied to distinguish the number of occurrences of the pattern, probability is the chance of appearing in the existing database, and utility occupancy is used to assess the contribution of selected patterns to the supporting transactions. The main contributions of this paper can be summarized as follows:

\begin{itemize}
	\item This paper presents an efficient UHUOPM algorithm aimed at discovering potential high utility-occupancy patterns from uncertain databases. To the best of our knowledge, this is the first study to address the problem of utility-driven exploiting high utility-occupancy patterns in uncertain data.
	
	\item To reduce the amount of access to the database, two list structures, named probability utility occupancy list (PUO-list) and probability frequency utility table (PFU-table), are constructed.
	
	\item Moreover, several pruning strategies are proposed to reduce the search space. A concept called the remaining utility occupancy is adopted to calculate the overestimated upper bound of patterns since the utility occupancy does not hold the downward closure property.
	
	\item To evaluate the performance of the compared algorithms, subsequent experiments were conducted on both real-life and synthetic datasets. The experiments show that several pruning strategies can effectively eliminate most of the unqualified patterns and improve the performance of the designed algorithm in terms of memory consumption and runtime.
\end{itemize}

The remainder of this paper is organized as follows: Related work is briefly introduced in Section \ref{sec:relatedwork}. To better explain the algorithm, some preliminaries are illustrated in Section \ref{sec:background}. In Section \ref{sec:algorithm}, the UHUOPM algorithm and several pruning strategies are introduced in detail. Furthermore, to verify the performance of the proposed algorithm, the experiments that were conducted are described in Section \ref{sec:experiments}. At last, a summary is given and future works are discussed in Section \ref{sec:conclusion}.

\section{Related Work}
\label{sec:relatedwork}

The related work consists of two areas, high-utility pattern mining and interesting pattern mining in uncertain databases. Details of the current developments and advances are presented below.

\subsection{High-utility pattern mining}

Data mining is a complex process of extracting and mining unknown and valuable patterns or laws from a large amount of data. Utility-driven mining is a branch of data mining, which focus on discovering mining patterns with high utility. In support-based pattern mining algorithms, information discovery merely needs to extract high-frequency patterns from the binary transaction database \cite{ahmed2009efficient, pei2001mining}. Here, whether a pattern appears in a transaction is a binary judgment. However, in real life, an item will appear more than once in a transaction and the frequency of the occurrence alone is not enough to measure how much utility a pattern brings to the supermarket or business company. Utility-driven pattern mining \cite{gan2018privacy,2gan2018survey,gan2018survey} combines external utility with local utility (i.e., the quantity of patterns in relative transactions) to calculate the utility of pattern. Provided that its overall utility value is greater than a predefined minimum utility threshold, the pattern is considered to be a high-utility pattern. Chan \textit{et al.} \cite{chan2003mining} designed a novel framework that considers not only the positive utility but also the negative utility to discover the top-\textit{k} high-utility patterns. Yao \textit{et al.} \cite{yao2004foundational} formally defined the concept of utility mining. Through this, profitable patterns can be found by combining external and internal utility. Moreover,  a mathematical model was proposed to predict the utility upper bounds of \textit{k}-itemset through a qualified \textit{k-1}-itemset. Liu \textit{et al.} \cite{liu2005two} then designed a transaction-weighted utilization model to unveil qualified patterns by taking advantage of transaction-weighted downward closure property to prune the inefficient patterns. Next, Liu \textit{et al.} \cite{liu2012mining} developed a more efficient algorithm named HUI-miner. This algorithm adopts the utility-list structure that contains the internal utility and remaining utility of the pattern in the supporting transactions. The upper bound of a pattern can be directly calculated by considering the information of its parent-node and parent-node's sibling nodes. If the utility of a pattern is less than the given defined minimum utility threshold, then the extension of this pattern can be directly pruned. Currently, the issues of HUIM has been extensively studied, such as ACO-based approach of HUIM \cite{wu2017aco}, mining high-utility association rules \cite{mai2017lattice}, HUIM with multiple minimum utility thresholds \cite{lin2016efficient2,krishnamoorthy2018efficient},  HUIM over data streams \cite{ryang2016high}, and so on.  While considering sequential data, the topic of high-utility sequential pattern mining has also been studied with methods like USpan \cite{yin2012uspan}, ProUM \cite{gan2019proum}, and HUSP-ULL \cite{gan2019fast}. Several studies of utility mining have been introduced to improve the mining effectiveness with constraints of various discount strategies \cite{lin2016fast} and discriminative patterns \cite{lin2017fdhup}. Developing effective and efficient algorithms for mining high-utility patterns is an active research area, and more recent studies can be referred to in the review literature \cite{gan2018survey}.

When the utility contribution ratio of a pattern is considered, the above-mentioned algorithms are not applicable. Tang \textit{et al.} \cite{tang2012incorporating} explained the concept of occupancy by introducing an application called investment portfolio recommendation and accordingly manifested a dominant and frequent itemset mining algorithm. Unfortunately, occupancy is merely based on the number of appearances of the pattern, and it cannot be applied to the range of utility. Subsequently, Shen \textit{et al.} \cite{shen2016ocean} proposed an OCEAN algorithm in which the utility occupancy is defined as the utility share of a pattern in supporting transactions. Nevertheless, this algorithm suffers from some shortcomings. Among them, the fatal disadvantage is that it can not discover the complete eligible patterns. To overcome that problem, Gan \textit{et al.} \cite{gan2019huopm} proposed two new data structures and the HUOPM algorithm. In HUOPM, the utility occupancy of the pattern $X$ in a supporting transaction is defined as the utility of pattern $X$ divided by the total utility in the transactions, and the utility occupancy of $X$ is the sum of each utility occupancy in all the supported transactions. This qualified pattern is said to be a high utility-occupancy pattern if the value obtained is no less than the minimum thresholds given.

\subsection{Interesting pattern mining in uncertain data}  

Due to sensor or network failure when collecting data in the real world, it is difficult to detect accurate or complete data. However, most algorithms have a preference for precise data and do not consider uncertain data. The above algorithms are all aimed at handling precise data. Therefore, it is essential to develop some algorithms to effectively discover useful patterns in uncertain databases \cite{aggarwal2009frequent}. Chui \textit{et al.} \cite{chui2007mining} first introduced a pioneering work to mine qualified frequent patterns in an uncertain database, and their Apriori-like UApriori algorithm adopts a hierarchical search measure, similar to the Apriori algorithm \cite{agrawal1994fast}, by comparing thresholds of support count and the probability to delete useless itemsets. The UApriori algorithm was optimized by Leung \textit{et al.} \cite{leung2008tree} using an extended frequent pattern tree structure. This method does not generate candidate itemsets and greatly improves the execution time and the mining performance. After that, Lin \textit{et al.} \cite{lin2012new} developed an algorithm based on a tree structure, called CUFF-tree to efficiently mine frequent patterns. In addition to mining frequent patterns in uncertain databases, it is also important to extract weighted frequent patterns \cite{gan2017extracting,lin2016weighted} or high-utility patterns \cite{lin2016efficient,lin2017efficiently} in uncertain databases. Lin \textit{et al.} \cite{lin2016efficient} proposed a novel framework, which is the potential high-utility itemset mining model. Several efficient algorithms, named PHUI-UP \cite{lin2016efficient}, PHUI-List \cite{lin2016efficient}, MUHUI \cite{lin2017efficiently},  CPHUI-List \cite{vo2020efficient}, and HUPNU \cite{gan2017mining}, were developed to find interesting patterns with both high utility and high probability. To summarize, the problem now is that according to what has been learned currently, 1) no work has utilized the concept of utility occupancy and uncertainty together to discover high utility-occupancy patterns in uncertain databases. 2) Besides, the measure of utility-occupancy is different from that of utility, in terms of definition, upper bound, and pruning strategies. Therefore, this paper is aimed at addressing this challenging task.

%%%%%%%%%%%%%%%%%%%%%%%%%%%%%%%%%%%%%%%%%%%
%%%%%%%%%%%%%%%%%%%% PRELIMINATIRES %%%%%%%%%%%%%%%
%%%%%%%%%%%%%%%%%%%%%%%%%%%%%%%%%%%%%%%%%%%

\section{Preliminary and Problem Statement} 
\label{sec:background}

To describe the proposed algorithm for finding potential high utility-occupancy patterns in the given databases, we used the uncertain database that is shown in Table \ref{table:db}, which is made up of ten transactions and five items that are distinct to each other. There are four parts to each entry: the transaction identifier, purchased items, number of relative items, and probability of each item. Let \textit{I} = \{\textit{i}$_{1}$, \textit{i}$_{2}$, $\ldots$, \textit{i$_{m}$}\} be a collection of items; and let \textit{D} = \{\textit{T}$_{1}$, \textit{T}$_{2}$, $\ldots$, \textit{T$_{n}$}\} be an uncertain quantitative database, where in supporting transactions, such as \textit{T}$_{k}$, each item \textit{i}$_{c}$ consists of three parts (the item name, the number of occurrences $q(\textit{i}_{c}, \textit{T}_{k})$ and the corresponding probability of occurrence $p(\textit{i}_{c}, \textit{T}_{k})$). The sum of the utility of each transaction is \textit{tu}. Table \ref{table:profit} shows the utility and profit of each item, which are manually defined. Table \ref{table:db} and Table \ref{table:profit} are taken as an example to explain the proposed algorithm below.

\begin{table}[!htbp]
	\newcommand{\tabincell}[2]{\begin{tabular}{@{}#1@{}}#2\end{tabular}}
	\centering
	\small
	\caption{Example of an uncertain quantitative database}
	\label{table:db}
	\begin{tabular}{|c|c|c|}
		\hline
		\textbf{\textit{tid}} & \textbf{Transaction (item, quantity, probability)} & \textbf{\textit{tu}} \\ \hline
		$ T_{1} $ & 	\{\textit{a}:2, 0.6\} \{\textit{c}:4, 0.8\} \{\textit{d}:7, 0.5\}  &  \$65 \\ \hline
		$ T_{2} $ & 	\{\textit{b}:2, 0.7\} \{\textit{c}:3, 0.4\}   & \$37 \\ \hline
		$ T_{3} $ &	    \{\textit{a}:3, 0.6\} \{\textit{b}:2, 0.6\} \{\textit{c}:1, 0.9\} \{\textit{d}:2, 0.8\}  &  \$38 \\ \hline
		$ T_{4} $ &	    \{\textit{b}:4, 0.5\} \{\textit{d}:3, 0.8\}  &  \$11 \\ \hline
		$ T_{5} $ &	    \tabincell{c}{\{\textit{a}:1, 0.9\} \{\textit{b}:3, 0.7\} \{\textit{c}:2, 0.9\} \\ \{\textit{d}:5, 0.6\} \{\textit{e}:1, 0.8\}} &   \$49  \\ \hline
		$ T_{6} $ &	    \{\textit{c}:2, 0.9\} \{\textit{e}:4, 0.8\}  &  \$58 \\ \hline
		$ T_{7} $ &	    \{\textit{c}:2, 0.4\} \{\textit{d}:1, 0.9\} &  \$23 \\ \hline
		$ T_{8} $ &	    \{\textit{a}:3, 0.6\} \{\textit{b}:1, 0.8\} \{\textit{d}:2, 0.8\} \{\textit{e}:4, 0.5\}  &  \$61 \\ \hline
		$ T_{9} $ &  	\{\textit{a}:2, 0.6\} \{\textit{c}:4, 0.5\} \{\textit{d}:1, 0.3\} &  \$59 \\ \hline
		$ T_{10} $ &	\{\textit{c}:3, 0.6\} \{\textit{e}:1, 0.7\} &  \$42 \\ \hline	
	\end{tabular}
\end{table}

\begin{table}[!htbp]
	\centering
	\small
	\caption{Unit utility of each item}
	\label{table:profit}
	\begin{tabular}{|c|c|}
		\hline
			\textbf{Item} &  \textbf{Utility (\$)} \\ \hline
		$ a $ & $ 7 $ \\ \hline
		$ b $ &	$ 2 $ \\ \hline
		$ c $ & $ 11 $ \\ \hline
		$ d $ & $ 1 $ \\ \hline
		$ e $ & $ 9 $ \\ \hline
	\end{tabular}
\end{table}

\begin{definition}[support count]
	\label{def_1}
	\rm Supposing that in a given database, several transactions contain itemset $X$, that is, itemset $X$ appears in these transactions, and then the number of occurrences is called \textit{SC} (support count) and denoted as \textit{SC}$(X)$ \cite{agrawal1994fast,han2004mining}. Then, the transactions that meet the conditions are put into a collection $\varGamma_X$, and thus the equation \textit{SC}$(X)$ = $|\varGamma_X|$ can be obtained. The pattern $X$ is considered a frequent pattern if and only if \textit{SC}$(X)$ is equal or greater than a predefined minimum support threshold $\alpha$.
\end{definition}

\begin{example} 
	In Table \ref{table:db}, it can be seen that pattern $(b)$ appears in transaction $T_2$, $T_3$, $T_4$, $T_5$, and $T_8$, respectively. Therefore, it can be concluded that \textit{SC}$(b)$ = 5. Similarly, \textit{SC}$(bc)$ = 3.
\end{example}
	
\begin{definition}[utility calculation]
	\label{def_2}
	\rm As shown in Table \ref{table:profit}, each item corresponds to a certain unit utility in the database. It represents the degree of preference of users for the product or the importance of the commodity as considered by experts. If $p(\textit{i}_{c})$ denotes the unit utility of each item, then the utility of the pattern $u(\textit{i}_{c}, \textit{T}_{k})$ = $p(\textit{i}_{c}) \times q(\textit{i}_{c}, \textit{T}_{k})$, where item $\textit{i}_{c}$ exists in the transaction $\textit{T}_{k}$. The utility of itemset $X$ in a supporting transaction can be expressed as $ u(X, T_{k})$ = $\sum _{i_{j}\in X \wedge X \subseteq T_{k}}u(i_{j}, T_{k}) $. Moreover, the utility of $X$ in a given database $D$ is defined as $u(X)$ = $\sum_{X\subseteq T_{k} \wedge T_{k}\in D} u(X, T_{k}) $. Finally, the transaction utility $(tu)$ is defined as the sum of the utility of all the items in this transaction.
\end{definition}

\begin{example}
	 For example, $u(b)$ = $u(b, T_2)$ + $u(b, T_3)$ + $u(b, T_4)$ + $u(b, T_5)$ + $u(b, T_8)$ = \$4 + \$4 + \$8 + \$6 + \$4 = \$26. $u(bc)$ = $u(bc, T_2)$ + $u(bc, T_3)$ + $u(bc, T_5)$ = \$37 + \$15 + \$28 = \$80. Thus, $tu(T_{1})$ = $u(a, T_1)$ + $u(c, T_1)$ + $u(d, T_1)$ = \$14 + \$44 + \$7 = \$65.
\end{example} 

\begin{definition}[utility occupancy \cite{gan2019huopm,shen2016ocean}]
	\label{def_3}
	\rm In a given transactional database, the utility contribution rate of an itemset in a database is also very significant, which is called the utility occupancy. The utility occupancy of an itemset $X$ in the relative supporting transaction $T_{k}$ is expressed as: 
	\begin{equation}
	uo(X, T_{k}) = \dfrac{u(X, T_k)}{tu(T_k)}. 
    \end{equation}
	Like the definition of utility, the calculation formula of utility occupancy of itemset $X$ in a database is defined as:
	 \begin{equation}
	 uo(X) = \dfrac{\sum_{X \subseteq T_k \wedge T_k \in D}uo(X,T_k)}{|\varGamma_X|},
	 \end{equation}
	 where $\varGamma_X$ is a collection of transactions containing itemset $X$, and $|\varGamma_X|$ is the length of the collection.
\end{definition}

\begin{example}
	For example, according to Definition \ref{def_2}, it can obtained that $tu(T_2)$ = \$37, $tu(T_3)$ = \$38, $tu(T_4)$ = \$11, $tu(T_5)$ = \$49, and $tu(T_8)$ = \$61. Therefore, it is simple to calculate $uo(b)$, which is first to compute the value of $\left.u(b, T_2) \middle/ tu(T_2)\right. + \left.u(b, T_3) \middle/ tu(T_3) \right. + \left.u(b, T_4) \middle/ tu(T_4)\right. + \left.u(b, T_5) \middle/ tu(T_5)\right. + \left.u(b, T_8) \middle/ tu(T_8)\right.$ and then divide this value by 5. Finally, it can be calculated that the final result is approximately 0.2257. Similarly, $uo(bc)$ can be calculated as 1.6553.
\end{example}

\begin{definition}
	\label{def_4}
	\rm This paper focuses on situations with uncertain databases, where the probability of data is used to represent the uncertainty. $pro(X)$ represents the probability of itemset $X$ in the corresponding transaction, and it can be denoted as $pro(X)$ =  $\sum_{i=1}^{|D|} \prod_{x_i \in X} p(x_i,T_k)$ \cite{lin2016efficient}. In the UHUOPM model, the possibility of a pattern is defined as $pro(X)$ = $\sum_{X \subseteq T_k \wedge T_k \in D} p(X, T_k)$.
\end{definition}

\begin{example}
	For example, $pro(b)$ = $p(b, T_2)$ + $p(b, T_3)$ + $p(b, T_4)$ + $p(b, T_5)$ + $p(b, T_8)$ = 0.7 + 0.6 + 0.5 + 0.7 + 0.8 = 3.3, $pro(bc)$ = $p(bc, T_2)$ + $p(bc, T_3)$ + $p(bc, T_5)$ + $p(bc, T_8)$ = 0.28 + 0.54 + 0.63 = 1.45.
\end{example}

\begin{definition}
	\label{def_PHUOP}
	\rm Given an uncertain database, if the support of an itemset $X$ is no less than the predefined minimum support threshold $ \alpha $, the utility occupancy is no less than the minimum utility occupancy threshold $ \beta $, and the probability is equal to or greater than the predefined minimum probability threshold $ \gamma $, then it can be called a potential high utility-occupancy pattern (\textit{PHUOP}). These thresholds are flexibly set according to the requirements of decision-makers based on their prior knowledge and interest.
\end{definition}

\begin{example}
	For example, based on the results obtained above, it is not difficult to obtain that \textit{SC}$(b)$ = 5, \textit{SC}$(bc)$ = 3, $uo(b)$ = 0.2257, $uo(bc)$ = 1.6553, $pro(b)$ = 3.3, and $pro(bc)$ = 1.45. Let $ \alpha $ be 0.3, $ \beta $ be 0.3, and $ \gamma $ be 0.05. After comparing each value with the corresponding threshold, we have that $uo(b)$ is less than $ \beta $. Thus, the itemset $b$ is not a PHUOP.
\end{example}

\begin{definition}
	\label{def_6}
	\rm Supposing that the items in the database are arranged in a certain order, such as alphabetical order or \textit{TWU}-ascending order, then there is no harm in reordering the items in the database in support of ascending order and expressing this order with the symbol $ \prec $.
\end{definition}

\begin{example}
	For example, in the above database, the support counts of each item can be easily obtained, and they are $SC(a)$: 5, $SC(b)$: 5, $SC(c)$: 8, $SC(d)$: 7, and $SC(e)$: 4. Since $SC(e) \leq SC(a) \leq SC(b) \leq  SC(d) \leq SC(c)$ holds, the support-ascending order is $e \prec a \prec b \prec d \prec c$.
\end{example}

\textbf{Problem Statement.}  The main goal of this paper is to first give an uncertain quantitative database and then discover interesting patterns that satisfy the conditions in which the support count is no less than the minimum support threshold $ \alpha $, the utility occupancy value is no less than the minimum utility occupancy threshold $ \beta $, and the probability value that exists in the database is equal or greater than the minimum probability threshold $ \gamma $. It is obvious that the task of mining PHUOPs depends upon three different parameters: $ \alpha $, $ \beta $, and $ \gamma $.

%%%%%%%%%%%%%%%%%%%%%%%%%%%%%%%%%%%%%%%%%%%%
%%%%%%%%%%%%%%%%%%%%% ALGORITHM %%%%%%%%%%%%%%%
%%%%%%%%%%%%%%%%%%%%%%%%%%%%%%%%%%%%%%%%%%%%
\section{Proposed Algorithm for Mining PHUOPs} 
\label{sec:algorithm}

In this section, two list-based structures, called probability utility occupancy list (PUO-list) and probability frequency utility table (PFU-table), are used respectively to store information in the database. They can reduce the execution time required for the database to be accessed and processed. Furthermore, by judging three elements, the support count, utility occupancy, and probability, the eligible patterns are selected.

\subsection{Two data structures}

Previous level-wise works of pattern mining (e.g., the well-known Apriori algorithm \cite{agrawal1994fast}) have adopted a hierarchical search strategy, and each level of pattern generation requires one access to the database, which greatly wastes memory space and increases execution time. The list-based HUIM algorithms (e.g., HUI-Miner \cite{liu2012mining} and HUOPM \cite{gan2019huopm}) creatively store the horizontal list structure and maintain the mining information required for discovering high-utility patterns. In this case, traversing the database many times is sufficient. Inspired by the idea of a vertical list structure (note that there are two common data structures - horizontal \cite{agrawal1994fast}  vs vertical \cite{zaki2003fast}), this paper proposes two list-based structures to store the necessary information for mining potential high utility-occupancy patterns. The two list-based structures are described in detail below.

\begin{definition}[remaining utility occupancy \cite{gan2019huopm,shen2016ocean}]
	\label{def_ruo}
	\rm Assume that all items in the database are sorted by $\prec$. The remaining utility occupancy (\textit{ruo})  of an itemset $X$ in a supporting transaction $T_k$ is defined as the sum of the utility occupancy of all items succeeding $X$ in this transaction and denoted as:
	\begin{equation}
	ruo(X, T_{k}) = \sum_{i \notin X \wedge X \prec i \wedge i \in T_k}uo(i, T_k).
	\end{equation}
	Let $\varGamma_X$ is a collection of transactions containing itemset $X$. The remaining utility occupancy of an itemset $X$ in a database $D$ is defined as: 
	\begin{equation}
	ruo(X) = \dfrac{\sum_{X \subseteq T_k \wedge T_k \in D}ruo(X,T_k)}{|\varGamma_X|}.
	\end{equation}
\end{definition}

\begin{definition}[PUO-list]
	\label{def_7}
	\rm Inspired by the UO-list \cite{gan2019huopm}, the probability utility occupancy list (PUO-list) is a collection of tuples where an itemset $X$ appears. It includes four elements $(\it{tid, pro, uo, ruo})$, which are an identifier of the transaction $(\it{tid})$, the probability value $(\it{pro})$, the utility occupancy value $(\it{uo})$, and the remaining utility occupancy value $(\it{ruo})$ of itemset $(\it{X})$. Among them, $(\it{pro})$ is the occurrence probability of itemset $(\it{X})$, $(\it{uo})$ is the proportion of utility in the transaction, and $(\it{ruo})$ in the given order database is the proportion of sum of the utility of all the items after itemset $(\it{X})$ in this transaction.
\end{definition}

\begin{figure}[!htb]
	\centering
 	\includegraphics[scale=0.63]{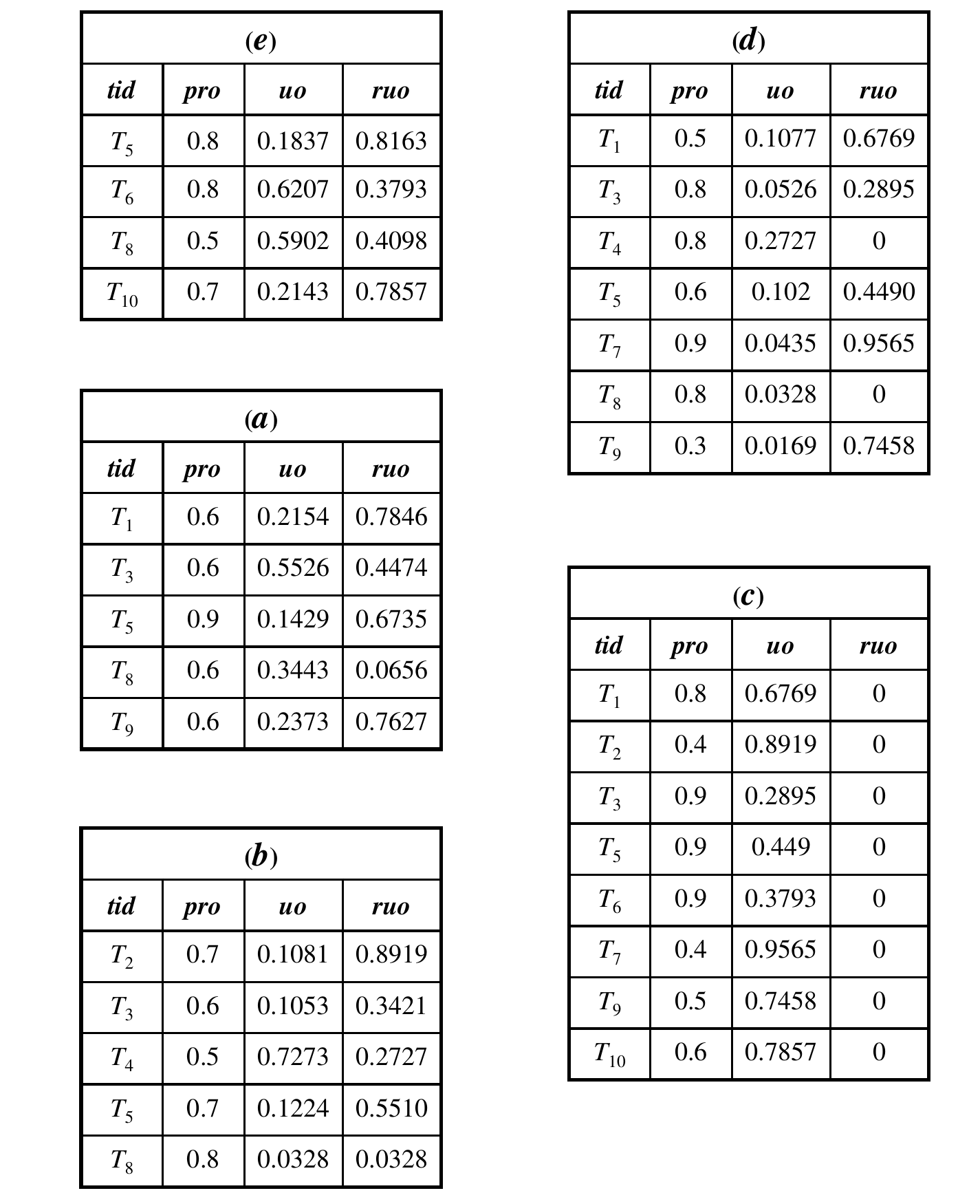}
	\caption{PUO-lists of five items}
	\label{fig:PUO-list}
\end{figure}

\begin{example}
	For example, each item in every transaction is first reordered in support count ascending order, as shown in the Table \ref{table:db1}. Then, a PUO-list using itemset $(e)$ as an example can be considered, where $(e)$ appears in transactions 5, 6, 8, and 10. The probability of $(e)$ is 0.8, the utility occupancy is 0.1837, and the remaining utility occupancy is 0.8163, which appears in transaction 5. Consequently, one tuple in the PUO-list of $(e)$ can be written as $(5, 0.8, 0.1837, 0.8163)$. The other tuples of $(e)$ are then calculated and listed in the same way. Finally, the PUO-lists of each item in Table \ref{table:db1} are listed in Fig. \ref{fig:PUO-list}.
\end{example}

\begin{table}[!htbp]
	\newcommand{\tabincell}[2]{\begin{tabular}{@{}#1@{}}#2\end{tabular}}
	\centering
	\small
	\caption{Revised uncertain quantitative database}
	\label{table:db1}
	\begin{tabular}{|c|c|c|c|}
		\hline
		\textbf{\textit{tid}} & \textbf{Transaction (item, quantity, probability)} & \textbf{\textit{tu}} \\ \hline
		$ T_{1} $ & 	\{\textit{a}:2, 0.6\} \{\textit{d}:7, 0.5\} \{\textit{c}:4, 0.8\}   &  \$65 \\ \hline
		$ T_{2} $ & 	\{\textit{b}:2, 0.7\} \{\textit{c}:3, 0.4\}   & \$37 \\ \hline
		$ T_{3} $ &	    \{\textit{a}:3, 0.6\} \{\textit{b}:2, 0.6\} \{\textit{d}:2, 0.8\} \{\textit{c}:1, 0.9\}   &  \$38 \\ \hline
		$ T_{4} $ &	    \{\textit{b}:4, 0.5\} \{\textit{d}:3, 0.8\}  &  \$11 \\ \hline
		$ T_{5} $ &	    \tabincell{c}{\{\textit{e}:1, 0.8\} \{\textit{a}:1, 0.9\} \{\textit{b}:3, 0.7\} \\ \{\textit{d}:5, 0.6\} \{\textit{c}:2, 0.9\}}   &   \$49  \\ \hline
		$ T_{6} $ &	    \{\textit{e}:4, 0.8\} \{\textit{c}:2, 0.9\}   &  \$58 \\ \hline
		$ T_{7} $ &	    \{\textit{d}:1, 0.9\} \{\textit{c}:2, 0.4\}  &  \$23 \\ \hline
		$ T_{8} $ &	    \{\textit{e}:4, 0.5\} \{\textit{a}:3, 0.6\} \{\textit{b}:1, 0.8\} \{\textit{d}:2, 0.8\}   &  \$61 \\ \hline
		$ T_{9} $ &  	\{\textit{a}:2, 0.6\} \{\textit{d}:1, 0.3\} \{\textit{c}:4, 0.5\}  &  \$59 \\ \hline
		$ T_{10} $	&	\{\textit{e}:1, 0.7\} \{\textit{c}:3, 0.6\}  &  \$42 \\ \hline	
	\end{tabular}
\end{table}

As shown in the designed PUO-list, it is easy to obtain the support count, the probability, and utility occupancy information of a target itemset in the entire processed database. For easier calculation, information is extracted from it and put into the PFU-table, as defined below.

\begin{definition}[PFU-table] 
	\label{def_8}
	\rm The information in probability frequency utility table (PFU-table) can be extracted from the PUO-list, including the itemset name, the number of its supporting transaction, the probability $(\it{pro})$, the utility occupancy $(\it{uo})$, and the remaining utility occupancy $(\it{ruo})$. Among them, the probability of an itemset $(\it{X})$ is the sum of probability in each transaction that contains it, and the average utility occupancy of an itemset $(\it{X})$ is equal to the average of effective utility occupancy in the corresponding PUO-list. Similarly, the average remaining utility occupancy is equal to the average of all remaining utility occupancy of  $(\it{X})$.
\end{definition}

\begin{figure}[!htb]
	\centering
 	\includegraphics[scale=0.6]{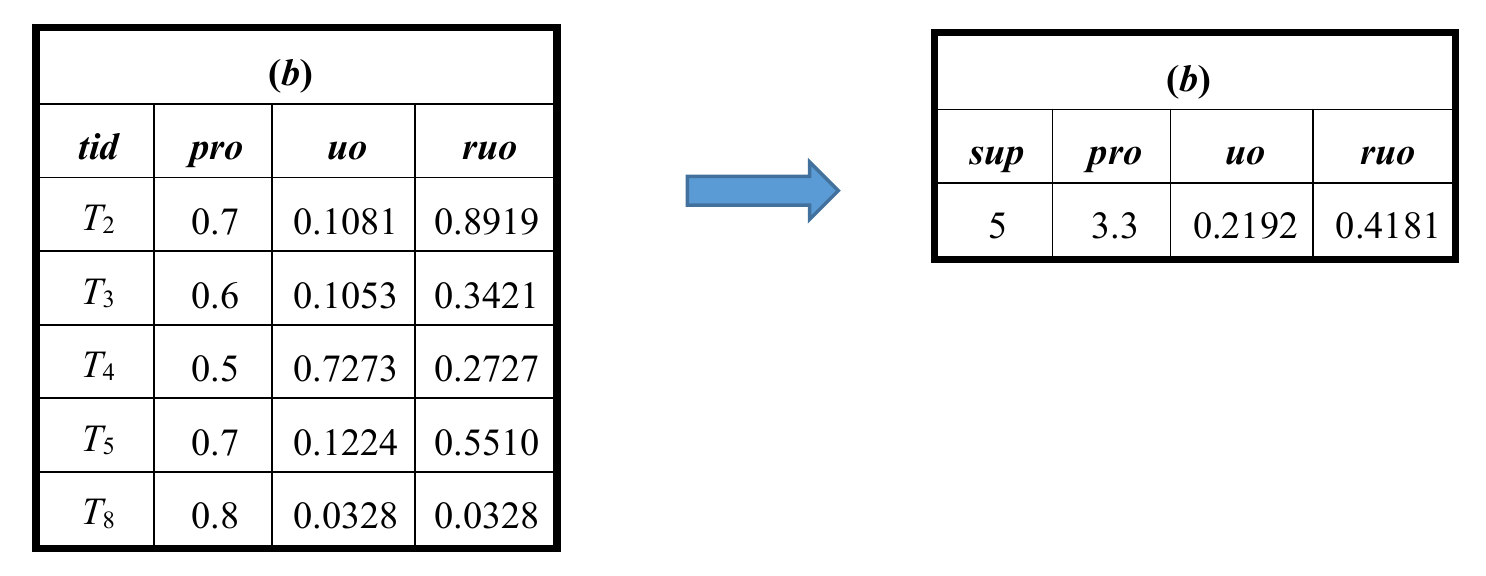}
	\caption{The PUO-list and PFU-table of itemset $(b)$}
	\label{fig:FWTableOfB}
\end{figure}

\begin{figure}[!htbp]
	\centering
 	\includegraphics[scale=0.63]{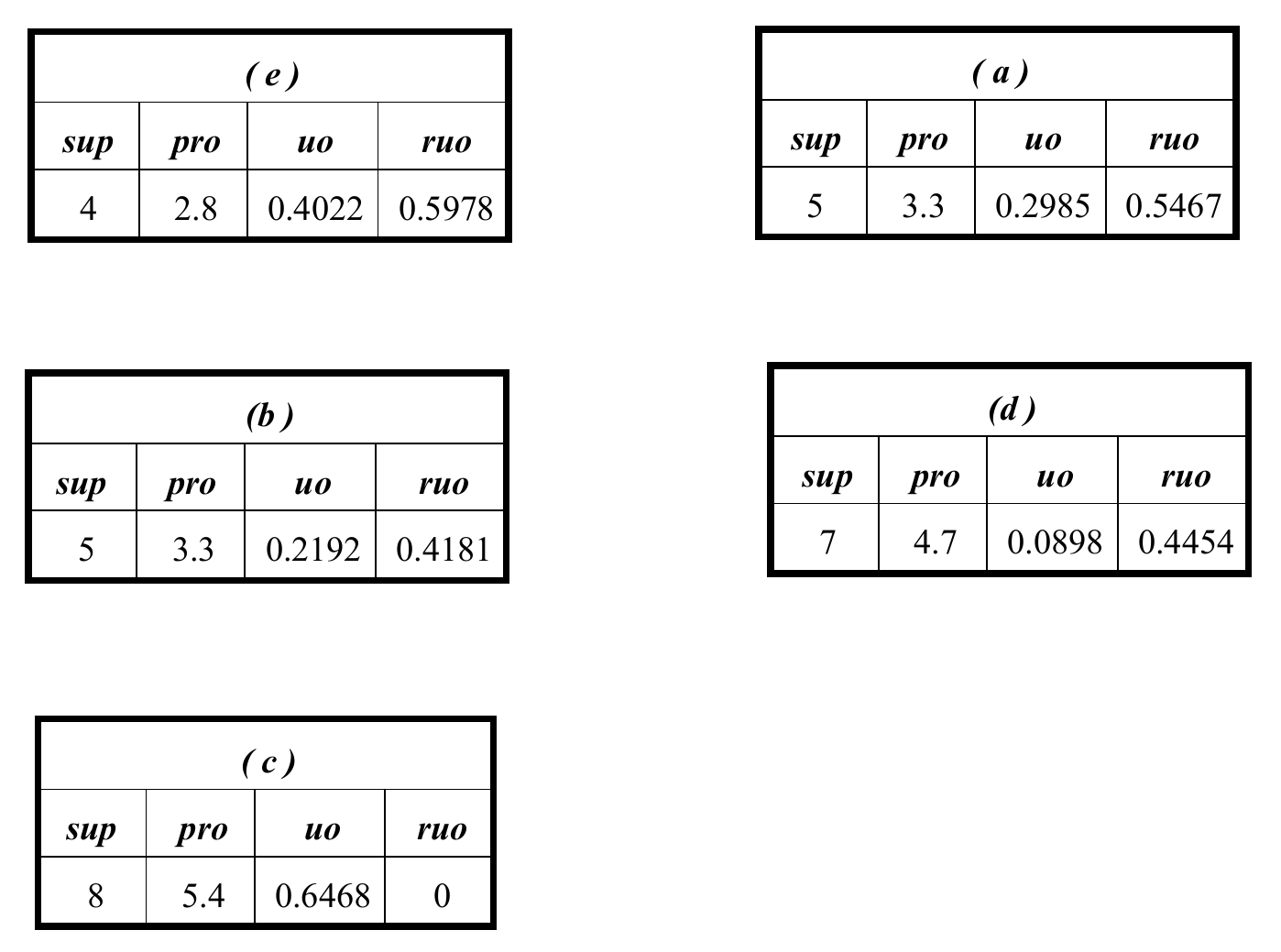}
	\caption{Constructed PFU-tables of all 1-itemsets}
	\label{fig:FP1}
\end{figure}

\begin{example}
  Using the PFU-table of an itemset $(b)$ as an example, its construction processes are presented below. Observing the PUO-list of itemset $(b)$ in Fig. \ref{fig:PUO-list}, it can be seen that $(b)$ appears in five transactions, and thus its support count is 5 and the sum of the probability of $(b)$ appearing in these five transactions is (0.7 + 0.6 + 0.5 + 0.7 + 0.8) = 3.3. The calculation process of its utility occupancy is (0.1081 + 0.1053 + 0.7273 + 0.1224 + 0.0328)/5 = 0.2192, and its remaining utility occupancy is (0.8919 + 0.3241 + 0.2727 + 0.5510 + 0328)/5 = 0.4181. Thus, the result of itemset $(b)$ is \{\textit{sup(b)}: 5, \textit{pro(b)}: 3.3, \textit{uo(b)}: 0.2192, \textit{ruo(b)}: 0.4181\}. The construction process of $(b)$ is shown in Fig. \ref{fig:FWTableOfB}, and the PFU-tables of all 1-itemsets in Table \ref{table:db1} are shown in Fig. \ref{fig:FP1}.
\end{example}

%%%%%%%%%%%     algorithm 1    %%%%%%%%%%%%%%%
\renewcommand{\algorithmicrequire}{\textbf{Input:}}%Input
\renewcommand{\algorithmicensure}{\textbf{Output:}}%Output
\begin{algorithm}[t]
	\label{construction}
	\caption{Construct($ X $, $ X_{a} $, $ X_{b} $)}
	\begin{algorithmic}[1]
		\REQUIRE $X$, an itemset  with its corresponding \textit{PUO-list} and \textit{PFU-table}; $ X_{a} $,  the extension of $X$ with an item $a$; $ X_{b} $, the extension of $X$ with an item $b$.
		\ENSURE	$ X_{ab} $.
		
		\STATE initialize  $ X_{ab}.PUO \leftarrow \emptyset $, $ X_{ab}.PFU\leftarrow \emptyset $; % $ X_{ab}.PFUT.sup:= 0 $,\;
		\STATE set \textit{supUB} = $X_{a}.PFU.sup$;
		\FOR {each tuple $ E_{a}\in X_{a}.PUO $}	
		\IF {$ \exists E_{a}\in X_{b}.PUO \wedge E_{a}.tid == E_{b}.tid $}
		\IF{$ X.PUO \neq \emptyset $}			
		\STATE search for $ E\in X.PUO, E.tid = E_{a}.tid $;			
		
		\STATE $E_{ab}$ $\leftarrow$ $<$$E_{a}.tid$, $E_{a}.pro \times E_{b}.pro / E.pro$, $E_{a}.uo$ + $E_{b}.uo$ - $E.uo$, $E_{b}.ruo$$>$;
		\STATE $ X_{ab}.PFU.pro $ += $E_{a}.pro \times E_{b}.pro / E.pro$;
		\STATE $ X_{ab}.PFU.uo $ += $ E_{a}.uo$ + $E_{b}.uo$ - $E.uo $;
		\STATE $ X_{ab}.PFU.ruo $ += $ E_{b}.ruo $;
		\ELSE
		
		\STATE $E_{ab}$ $\leftarrow$ $<$$E_{a}.tid$, $E_{a}.pro \times E_{b}.pro$, $E_{a}.uo$ + $E_{b}.uo$, $E_{b}.ruo$$>$;
		\STATE $ X_{ab}.PFU.pro $ += $ E_{a}.pro$ $\times E_{b}.pro $;
		\STATE $ X_{ab}.PFUT.uo $ += $ E_{a}.uo$ + $E_{b}.uo $;
		\STATE $ X_{ab}.PFU.ruo $ += $ E_{b}.ruo $;
		\ENDIF
		\STATE $ X_{ab}.PUO \leftarrow X_{ab}.PUO \cup E_{ab} $;
		\STATE $  X_{ab}.PFU.sup $ ++;	
		\ELSE		
		\STATE \textit{supUB} - -;
		\IF{\textit{supUB} $< \alpha \times |D| $}
		\STATE \textbf{return} \textit{null};	
		\ENDIF
		\ENDIF
		\ENDFOR
		\STATE \textbf{return} $ X_{ab} $\
	\end{algorithmic}
\end{algorithm}
%%%%%%%%%%%     algorithm 1    %%%%%%%%%%%%%%%

When the PUO-lists and PFU-tables of the 1-itemsets have been constructed, there is no need to follow these processes to build them for $k$-itemsets $(k > 1)$ by rescanning the database. Instead of traversing the database multiple times, the following construction based on the PUO-list and PFU-table of the 1-itemsets is used, where the required information is already contained. Algorithm 1 shows how to calculate $k$-itemsets $(k \geq 2)$ based on 1-itemsets. At the beginning, an itemset $X$ and two extensions of it, namely, $X_a$ and $X_b$, are given, and the order of $a$ precedes $b$. A new itemset $X_{ab}$ can be obtained by combining these two extensions. The algorithm also involves a pruning strategy, which is explained in detail in the next subsection. In Algorithm 1, lines 5 to 16 illustrate two cases of whether $X$ is an empty set. If $X$ is an empty set (lines 12 to 15), then the probability of $X_{ab}$ is directly multiplied by the probability of $X_a$ and $X_b$ appearing in the same transaction. Moreover, its utility occupancy is the sum of utility occupancy of $X_a$ and $X_b$, and the remaining utility occupancy is the same as in the later itemset w.r.t. the total order. If $X$ is not an empty set (lines 5 to 10), then the probability of $X_{ab}$ should be the probability of $X_a$ multiplied by the probability of $X_b$ and then divided by the probability of $X$. Besides, the utility occupancy of $X_{ab}$ equals $X_a$ plus $X_b$ and then subtracts that of $X$.

\subsection{Upper bound on probability and utility occupancy}

It is widely known that the Apriori algorithm \cite{agrawal1994fast} has the downward closure property of support, which means that if a $k$-itemset is a frequent pattern, then any of its subsets should be frequent. On the contrary, if a $k$-itemset is not a frequent pattern, then its superset should be not frequent either. Making use of this property can greatly reduce the search space and the execution time in those support-based pattern mining models. However, this property is not applicable for utility occupancy. For example, when the minimum threshold for utility occupancy is set to 0.3, the utility occupancy value of itemset $(ab)$ is 0.4334, which is a high utility-occupancy itemset. Besides, that value of $a$ is 0.2985, which does not satisfy the requirements. In general, constructing the two structures of all the itemsets requires much memory and runtime, which is quite expensive. Consequently, it is necessary to find the upper bound on utility occupancy, which is called $\hat{\phi}$.

\begin{figure}[!htbp]
	\centering
 	\includegraphics[scale=0.65]{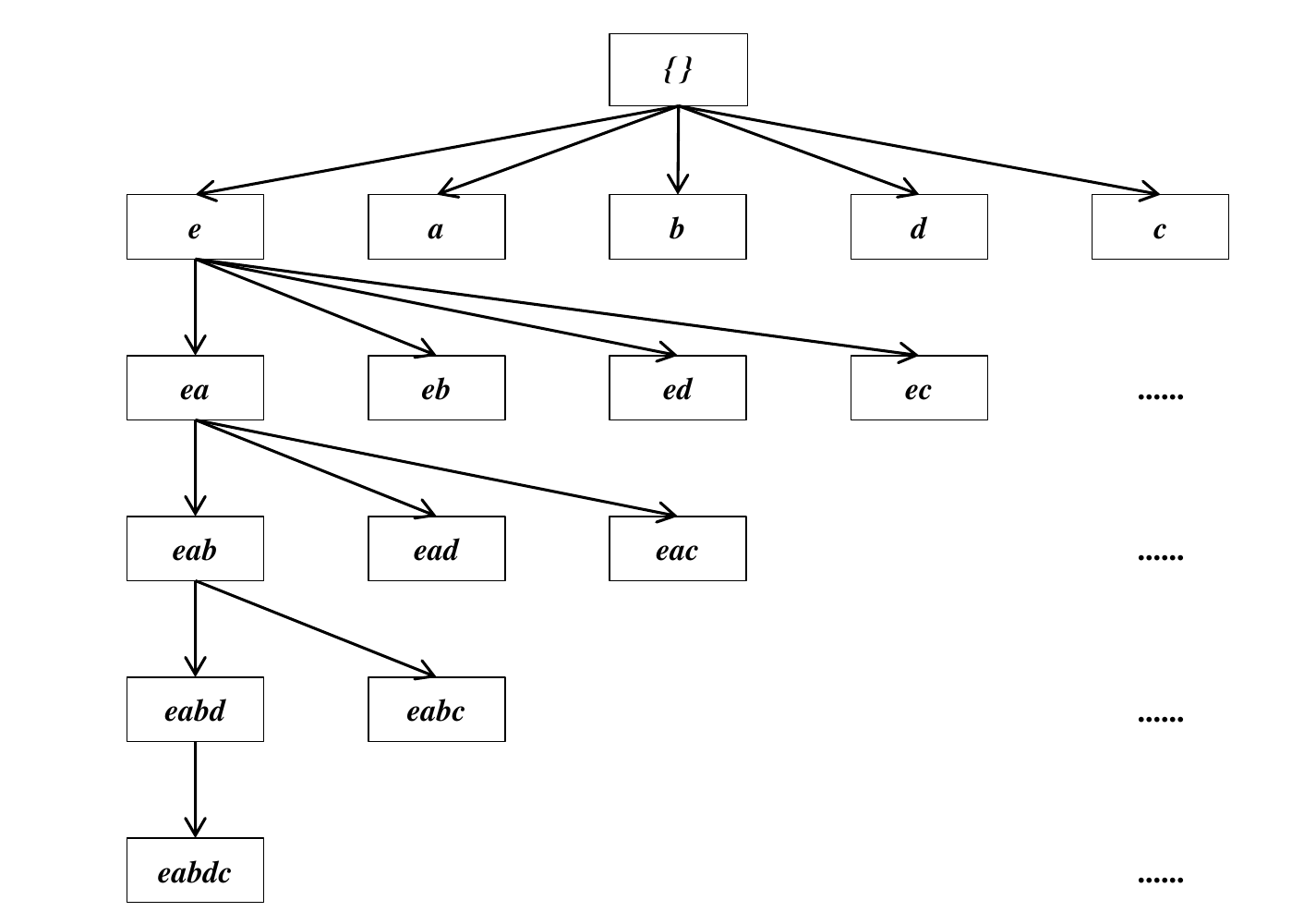}
	\caption{SC-tree} 
	\label{fig:tree}
\end{figure}

\begin{definition}[SC-tree] 
	\label{def_9} 
	\rm  According to a previous study \cite{rymon1992search}, a set-enumeration tree can be constructed and enumerated in a certain order. In the UHUOPM algorithm, the order of the ascending support count is taken as the overall order of the set-enumeration tree, and the full name is Support-Count tree (SC-tree). A part of this specific tree is shown in Fig. \ref{fig:tree}.
\end{definition}

\begin{lemma}
	\label{lemma_upperbound}
	For any node $X$ in the SC-tree, the upper bound on the utility occupancy of its child node $Y$ can be calculated as $\phi$ = $	\dfrac{\sum_{top \alpha \times |D|, T_k \in \varGamma_X}\{uo(X,T_k) + ruo(X,T_k)\}^{\downarrow}}{|\alpha \times |D||}$ \cite{gan2019huopm}, where $|D|$ denotes the number of transactions in the database and $\varGamma_X$ is the collection of transactions that contain itemset $X$. Besides, $top$ and $\downarrow$ signify that the values of the utility occupancy are sorted in descending order, and the top $\alpha \times |D|$ values are utilized for further calculation, in which $k$ is the length of $X$ (a $k$-itemset). The detailed proof of this upper bound can be referred to in prior work \cite{gan2019huopm}. 
\end{lemma}

\begin{example}
	For example, consider the node $c$ and its two subsets, $ca$ and $cd$, in the SC-tree. The utility occupancy of $c$ can be calculated by Definition \ref{def_3} and the value is 0.6468. Similarly, according to Lemma \ref{lemma_upperbound}, the upper bound on the subset with a size of 2 with $c$ as the root node can be calculated as 0.3081. This upper bound is greater than the threshold of utility occupancy.
\end{example}

\begin{lemma}
	\label{lemma_pro}
	Suppose there exists itemset $X_k$ (containing $k$ items) and $X_{k-1}$ (containing $k$-1 items) in an uncertain database, and $X_{k-1}$ is a subset of $X_k$. If $X_k$ is a high probability itemset, then $X_{k-1}$ should be a high probability itemset too. In other words, the high probability itemset has a downward closure property, such as $pro(X_k)$ $\textless$ $pro(X_{k-1})$ \cite{lin2016efficient}.
\end{lemma}

\begin{example}
	For example, the probability of node $c$ appearing in the running example is 5.4 while that of node $ca$ is 2.13. The former should be greater than or equal to the latter.
\end{example}

\subsection{Proposed algorithm and pruning strategy}
This section describes the proposed algorithm and the effective pruning strategies in detail. Given the several parameters involved, the utilized pruning strategies are mainly based on support count, probability, and utility occupancy. The adopted strategies are presented below.

\begin{strategy}
	\label{stra_1}
	When depth-first traversing the designed SC-tree as mentioned above, if the support count of a node $X$ is less than the user-defined minimum support threshold $\alpha$ multiplied by the database size, then this node and its descendants can be directly pruned.
\end{strategy}

\begin{proof}
	\rm  This strategy is based on the Apriori algorithm \cite{agrawal1994fast}, and the property can be extracted as \textit{SC}$(X_k)$ $\leq$ \textit{SC}$(X_{k-1})$. There is no doubt that if \textit{SC}$(X_{k-1}) < \alpha \times |D|$, then \textit{SC}$(X_k)$ $\textless$ $\alpha \times |D|$ and $X_k$ can be directly pruned.
\end{proof}

\begin{strategy}
	\label{stra_2}
	In an SC-tree constructed in $\prec$ order, if the upper bound on utility occupancy of their offspring is calculated based on a node $X$, which is less than the user-defined minimum threshold $\beta$, then all the nodes rooted at $X$ as descendant nodes can be quickly pruned.
\end{strategy}

\begin{proof}
	\rm  After building the corresponding list structures for a tree node $X$ in SC-tree, the upper bound on utility occupancy of $X$ can be quickly calculated using Lemma \ref{lemma_upperbound}. Since this value is derived from the utility occupancy and the remaining utility occupancy, if the upper bound is less than the minimum threshold $\beta$, then there is no need to build the PUO-lists of any descendant nodes of $X$.
\end{proof}

\begin{strategy}
	\label{stra_3}
	In the designed SC-tree, if the overall probability of a pattern $X$ is greater than or equal to the minimum probability threshold $\gamma$, then this pattern is a high probability pattern. On the contrary, if the value is less than the threshold, then node $X$ and all nodes with it as the root will be pruned.
\end{strategy}

\begin{proof}
	\rm  Like Strategy \ref{stra_1}, based on Lemma \ref{lemma_pro}, we can obtain $pro(X_k)$ $\textless$ $pro(X_{k-1})$. It is easy to acquire that $pro(X_{k})$ $\textless$ $\gamma$ in the case of $pro(X_{k-1})$ $\textless$ $\gamma$.
\end{proof}

\begin{strategy}
	\label{stra_4}
	One step closer to Strategy \ref{stra_1} in Algorithm \ref{construction}, if the support count that itemset $X_a$ holds is less than or equal to the minimum support threshold $\alpha \times |D|$, then it is not necessary to measure its extended itemset $X_{ab}$.
\end{strategy}

\begin{proof}
	\rm  The function of Strategy \ref{stra_4} is the same as Strategy \ref{stra_1}, except that Strategy \ref{stra_4} strengthens the judgment at the end of the proposed UHUOPM algorithm.
\end{proof}

Feasible strategies for trimming the search space and reducing the runtime are proposed above. The core processes of the UHUOPM algorithm are explained according to these proposed strategies and shown below.

\renewcommand{\algorithmicrequire}{\textbf{Input:}}%Input
\renewcommand{\algorithmicensure}{\textbf{Output:}}%Output
\begin{algorithm}[t]
	\label{UHUOPM-algorithm}
	\caption{UHUOPM (\textit{D}, \textit{utable}, $\alpha$, $\beta$, $\gamma$)}
	\begin{algorithmic}[1]	
		\REQUIRE an uncertain transaction database $D$, utility table $utable$, the minimum support threshold $ \alpha $, the minimum utility occupancy threshold $\beta$, and the minimum probability threshold $\gamma$.
		\ENSURE	\textit{PHUOPs}.	
			
		\STATE scan $D$ to calculate the \textit{SC}$(i)$ and $pro(i)$ of each item $ i \in I $ and the $tu$ value of each transaction;
		\STATE find $ I^* \gets \left\{  i \in I | SC(i) \geq \alpha \times |D| \land pro(i)) \geq \gamma \times |D|  \right\} $;
		
		\STATE sort $ I^* $ in the designed total order $ \prec $, such as ascending support count;
		\STATE using the total order $ \prec $, scan $D$ once to build the PUO-list and PFU-table for each 1-item $ i\in I^*$;
		\STATE \textbf{call \textit{PHUOP-Search}}($\phi, I^*, \alpha, \beta ,\gamma$).		
		\STATE \textbf{return} \textit{PHUOPs}\
	\end{algorithmic}
\end{algorithm}

For the UHUOPM algorithm, the processed database with its utility-table and three parameters are needed in advance. They are the uncertain quantitative database $D$, the unit utility corresponding to each item w.r.t. \textit{utable}, the minimum support threshold $\alpha$, the minimum utility occupancy threshold $\beta$, and the minimum probability threshold $\gamma$. At the beginning of Algorithm 2, during the first traversal of the database, the support count and corresponding probability for each item are calculated. At the same time, the transaction utility (\textit{tu}) of the transactions are calculated according to Definition \ref{def_2}, which will be used in the subsequent processes. Then, the 1-itemsets $I^*$ whose support count and probability meet the conditions are filtered out, and these itemsets in every processed transaction in the ascending order of their support counts are reordered. After that, the database is traversed again to construct the corresponding PUO-lists and PFU-table for each 1-itemset in $I^*$. After the initial processes, the next step is to filter out PHUOPs based on the given conditions. More details are given in Algorithm 3.

\renewcommand{\algorithmicrequire}{\textbf{Input:}}%Input
\renewcommand{\algorithmicensure}{\textbf{Output:}}%Output
\begin{algorithm}[t]
	\label{PHUOP-Search procedure}
	\caption{PHUOP-Search ($X$, $\textit{extenOfX}$, $\alpha$, $\beta$, $\gamma$)}
	\begin{algorithmic}[1]			
		\REQUIRE	an uncertain transaction database $D$, an itemset $X$ and its extended itemsets \textit{extenOfX}, the minimum support threshold $ \alpha $, the minimum utility occupancy threshold $\beta$, and the minimum probability threshold $\gamma$.
		\ENSURE	 \textit{PHUOPs}.

		\FOR {each itemset $ X_{a}\in $ $ \textit{extenOfX} $}	
		\STATE obtain $ SC(X_a) $ and $ uo(X_a) $  from the built $ X_{a}.PFU $;
		\IF{$ SC(X_a) \geq \alpha \times |D| \land pro(X_a) \geq \gamma \times |D|$}
		
		\IF{$ uo(X_a)\geq \beta $}			 
		\STATE \textit{PHUOPs} $ \leftarrow$ \textit{PHUOPs} $\cup X_{a} $;	
		\ENDIF
		
		\STATE	$ \hat{\phi}(X_a) \leftarrow  \textbf{\textit{UpperBound}}(X_a.PUO, \alpha) $;
		\IF{$ \hat{\phi}(X_a) \geq \beta $}
		
		\STATE $ \textit{extenOfX}_{a}\leftarrow  \emptyset $;
		\FOR {each $ X_{b}\in \textit{extenOfX} $ that $ X_{a} $   $ \prec $  $ X_{b} $}
		
		\STATE $ X_{ab}\leftarrow X_{a} \cup X_{b} $;
		\STATE call $ \textbf{\textit{Construct}}(X, X_{a}, X_{b}) $;
		
		\IF{$ X_{ab}.PUO \not= \emptyset$}
		\IF{$SC(X_{ab}) \geq \alpha \times |D| \land pro(X_{ab}) \geq \gamma \times |D|$}  
		
		\STATE $ \textit{extenOfX}_{a}\leftarrow \textit{extenOfX}_{a}\cup X_{ab}.PUO $;	
		\ENDIF	
		\ENDIF	
     			
		\ENDFOR		 						  		
		\STATE \textbf{call \textit{PHUOP-Search}}$\boldmath{(X_{a}, \textit{extenOfX}_{a}, \alpha, \beta, \gamma)}$;	
		\ENDIF
		\ENDIF
		\ENDFOR
		\STATE \textbf{return} \textit{PHUOPs}\
	\end{algorithmic}
\end{algorithm}

For Algorithm 3, the input consists of a prefix itemset $X$, which is initially a set of extended itemsets \textit{extendOfX} that is composed of the combination of $X$ and each of items after it and the three user-specified thresholds that are used as judgment conditions. The algorithm mainly adopts recursion to reduce the amount of computation and performs a depth-first traversal on the SC-tree. For each itemset $X_a$ in the extension of $X$, it is easy to obtain its support count and probability from the corresponding PFU-table. If the values of these two parameters meet the conditions, then this itemset can participate in the subsequent calculation. Next, the utility occupancy of $X_a$ is calculated and if it is greater than $\gamma$, then, according to the previous definitions, this itemset is the PHUOP that we want to discover. On the contrary, if it does not meet the conditions of utility occupancy threshold, then the upper bound $\hat{\phi}(X_a)$ of this itemset extension \textit{extendOfX} is calculated and it is assumed that this upper bound is greater than $\beta$, which means that \textit{extendOfX}$_a$ may be a PHUOP. In the next step, each itemset $X_{k-1}$ in \textit{extendOfX}$_a$ is combined with the itemset after itself to form $X_{k}$ and  two lists are accordingly constructed. The specific construction process can be referred to in Algorithm 1. If the newly constructed itemset meets the two basic conditions of PHUOP w.r.t. support count and probability, then this itemset can be put into the set for subsequent iterative processes.

Strategy \ref{stra_4} is applied in lines 20 to 22 in Algorithm 1. The support count of the extension of $X_a$ should be equal to or less than that of $X_a$. Using this condition, whether the extension of $X_a$ can be directly trimmed is determined without calculating other conditions. Algorithm 4 develops the design upper bound calculation formula obtained by Lemma \ref{lemma_upperbound}. The entire algorithm adopts a pruning strategy, which can efficiently prune some unpromising nodes in the SC-tree.

%%%%%%%%%%%%%%%%%   Algorithm 4   %%%%%%%%%%%%%%%%%%%
\renewcommand{\algorithmicrequire}{\textbf{Input:}}%Input
\renewcommand{\algorithmicensure}{\textbf{Output:}}%Output
\begin{algorithm}[t]
	\label{UpperBound}
	\caption{UpperBound ($ X_a.PUO $, $ \alpha $)}
	\begin{algorithmic}[1]	
		\REQUIRE an uncertain transaction database $D$, itemset $X_a$ and its corresponding PUO-list, the minimum support threshold $ \alpha $.
		
		\ENSURE	the upper bound on $X_a$, $ \hat{\phi}(X_a) $.
						
		\STATE  \textit{sumTopK} $ \leftarrow 0, \hat{\phi}(X_a) \leftarrow 0, V_{occu} \leftarrow \emptyset $;
		\STATE calculate $ (uo(X,T_k) + ruo(X,T_k)) $ of each tuple from the built $ X_{a}.PUO $ and put them into the set of $ V_{occu} $;
		\STATE sort $ V_{occu} $ by descending order as $ V_{occu}^{\downarrow} $;
		
		\FOR {$ k \leftarrow $ 1 to  $\alpha \times |D| $ in $ V_{occu}^{\downarrow} $}		
		\STATE   \textit{sumTopK}  $\leftarrow $  \textit{sumTopK} +  $V_{occu}^{\downarrow}[k]  $;					
		\ENDFOR
		
		\STATE $ \hat{\phi}(X_a) = \dfrac{sumTopK}{\alpha \times |D|} $.		
		\STATE \textbf{return} $ \hat{\phi}(X_a) $
	\end{algorithmic}
\end{algorithm}
%%%%%%%%%%%%%%%%%   Algorithm 4   %%%%%%%%%%%%%%%%%%%

%%%%%%%%%%%%%%%%%%%%%%%%%%%%%%%%%%%%%%%%%%%%
%%%%%%%%%%%%%%%%%%%%% EXPERIMENT %%%%%%%%%%%%%%%
%%%%%%%%%%%%%%%%%%%%%%%%%%%%%%%%%%%%%%%%%%%%
\section{Experiments} 
\label{sec:experiments}
This section describes the experiments that were conducted. The experimental results can be used to determine whether the performances of the compared algorithms were acceptable (both efficient and effective) or not. It should be noted that this is the first paper that combines utility occupancy and uncertainty embedding in databases to utility-driven discover potential high utility-occupancy patterns. The OCEAN \cite{shen2016ocean}  and HUOPM \cite{gan2019huopm} methods are closely related to the current research work. OCEAN is the first algorithm for mining HUOPs, while it cannot discover the complete final results. Thus, OCEAN is not compared as the baseline to evaluation the proposed model, and HUOPM is the best existing algorithm for mining utility-occupancy patterns. In the recent literature, several utility mining methods, e.g., PHUI-UP \cite{lin2016efficient}, PHUI-List \cite{lin2016efficient}, CPHUI-List \cite{vo2020efficient}, and MUHUI \cite{lin2017efficiently},  have been proposed to deal with uncertain databases. However, all these methods do not measure the concept of utility-occupancy. Utility-driven mining aims to explore the interesting patterns by taking utility into account. In addition, utility and utility-occupancy are two different measures, as mentioned before.

Therefore, to evaluate whether the proposed algorithm is acceptable, the proposed UHUOPM algorithm was compared with the state-of-the-art HUOPM algorithm in terms of runtime, visited nodes, and the number of derived patterns. Since high utility-occupancy patterns mining is a relatively novel concept, there is no other comparison algorithm that can be used in experiments for evaluation the proposed model. We also included two variants in the comparison to further evaluate the performance of the proposed pruning strategies. We call these two algorithms UHUOPM$_1$ (using the pruning strategies 1 and 2) and UHUOPM$_2$ (using the pruning strategies 1 and 3).

\subsection{Experimental setup and datasets}

All the experimental procedures were written in Java and the program was ran on a desktop computer. The computer's basic configuration included 4GB of memory and 64-bit Windows 7 operating system.

To better evaluate the performance of the compared algorithms, this experiment involved not only realistic datasets but also artificially synthesized datasets. Both real-life datasets (consisting of retail, mushroom, and kosarak) and the synthesized dataset (T40I10D100K) were selected for the experiments. These datasets included sparse, dense, short, and long features, and the algorithms could be compared in a comprehensive manner. Among them, the data source of the retail dataset consists the sales of a retail store in Belgium, which is a sparse dataset. The mushroom dataset aims to determine whether it is a poisonous mushroom by judging the 22 characteristics of a mushroom, which is a dense dataset. The kosarak dataset is longer than the other two datasets. T40I10D100K is a synthetic dataset. Sparse means that the number of items in the dataset is small, the length of the items is short and the dataset contains few transactions while a dense dataset is the opposite. The main features of these datasets are described in detail in Table \ref{table:features}. In this table, $|D|$ represents the number of transactions and $|I|$ implies the number of distinct items contained in the dataset.

\begin{table}[!htbp]
	\newcommand{\tabincell}[2]{\begin{tabular}{@{}#1@{}}#2\end{tabular}}
	\centering
	\small
	\caption{Features of the datasets}
	\label{table:features}
	\begin{tabular}{|c|c|c|c|}
		\hline
		\textbf{\textit{Datasets}} & \textbf{\textit{$|D|$}} & \textbf{\textit{$|I|$}} & \textbf{\textit{Type}}\\ \hline
		mushroom & 	8,124  &  120 &  dense \\ \hline
		retail & 	88,162  &  16,470 &  sparse \\ \hline
		kosarak & 	990,002  &  41,271 &  sparse \\ \hline
		T40I1D100K & 	100,000  &  1000 &  sparse \\ \hline
		
	\end{tabular}
\end{table}

\subsection{Runtime analysis}

The runtime of the four algorithms is evaluated below. For comparison of two factors (utility occupancy and uncertainty), one is assumed to be fixed and the other one is set differently. Since the algorithm involves in three parameters, experiments on the three ingredients needed to be performed separately. For example, when comparing the effects of utility occupancy, we needed to set the minimum thresholds of support and probability to be constant. It should be noted that the minimum support threshold is referred to as \textit{MS}, the minimum utility occupancy threshold is referred to as \textit{MUO}, and the minimum probability threshold is referred to as \textit{MP}.

% % % % % % % % % % % % % % %
\begin{figure}[!hbt]
	\centering %所给出的四个数字 分别代表了从左、下、右、上被截去的值
 	\includegraphics[trim=50 0 30 0,clip,scale=0.5]{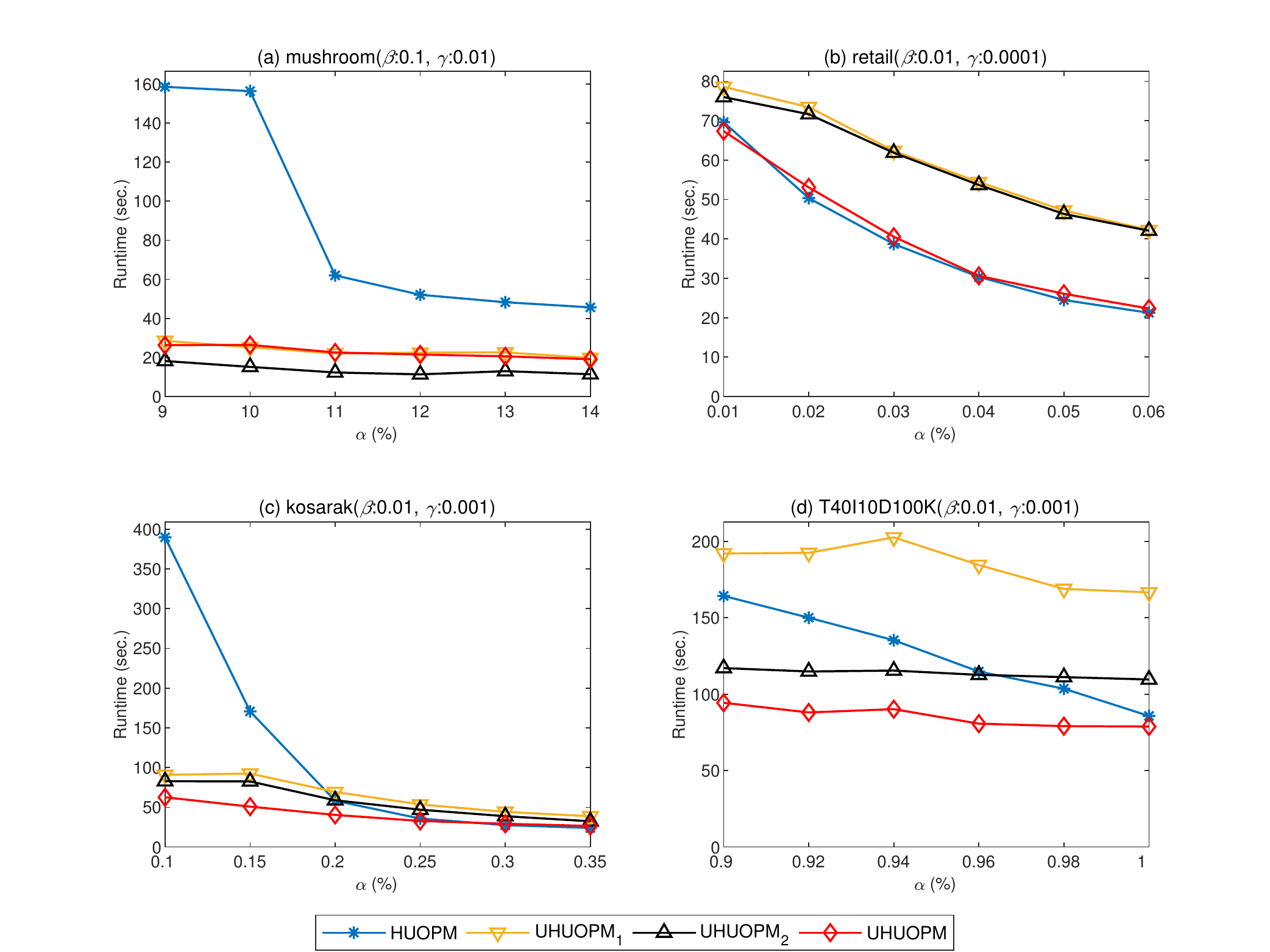}
	\captionsetup{justification=centering}
	\caption{Runtime under a changed $\alpha$ with a fixed $\beta$ and $\gamma$}
	\label{fig:runtimeMS}	
\end{figure}
% % % % % % % % % % % % % % %

% % % % % % % % % % % % % % %
\begin{figure}[!hbt]
	\centering %所给出的四个数字 分别代表了从左、下、右、上被截去的值
 	\includegraphics[trim=50 0 30 0,clip,scale=0.5]{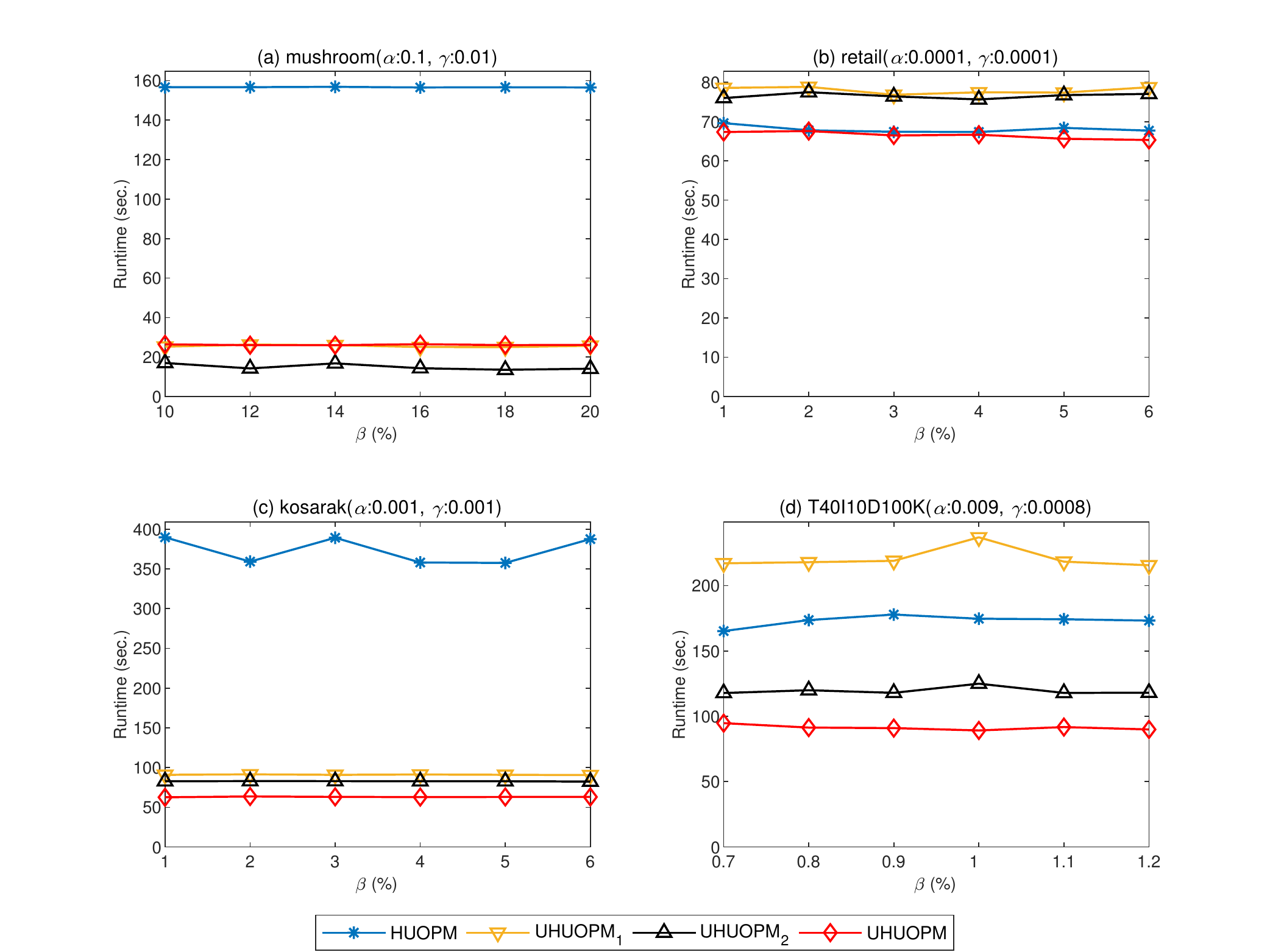}
	\captionsetup{justification=centering}
	\caption{Runtime under a changed $\beta$ with a fixed $\alpha$ and $\gamma$}
	\label{fig:runtimeMU}	
\end{figure}
% % % % % % % % % % % % % % %

% % % % % % % % % % % % % % %
\begin{figure}[!hbt]
	\centering %所给出的四个数字 分别代表了从左、下、右、上被截去的值
 	\includegraphics[trim=50 0 30 0,clip,scale=0.5]{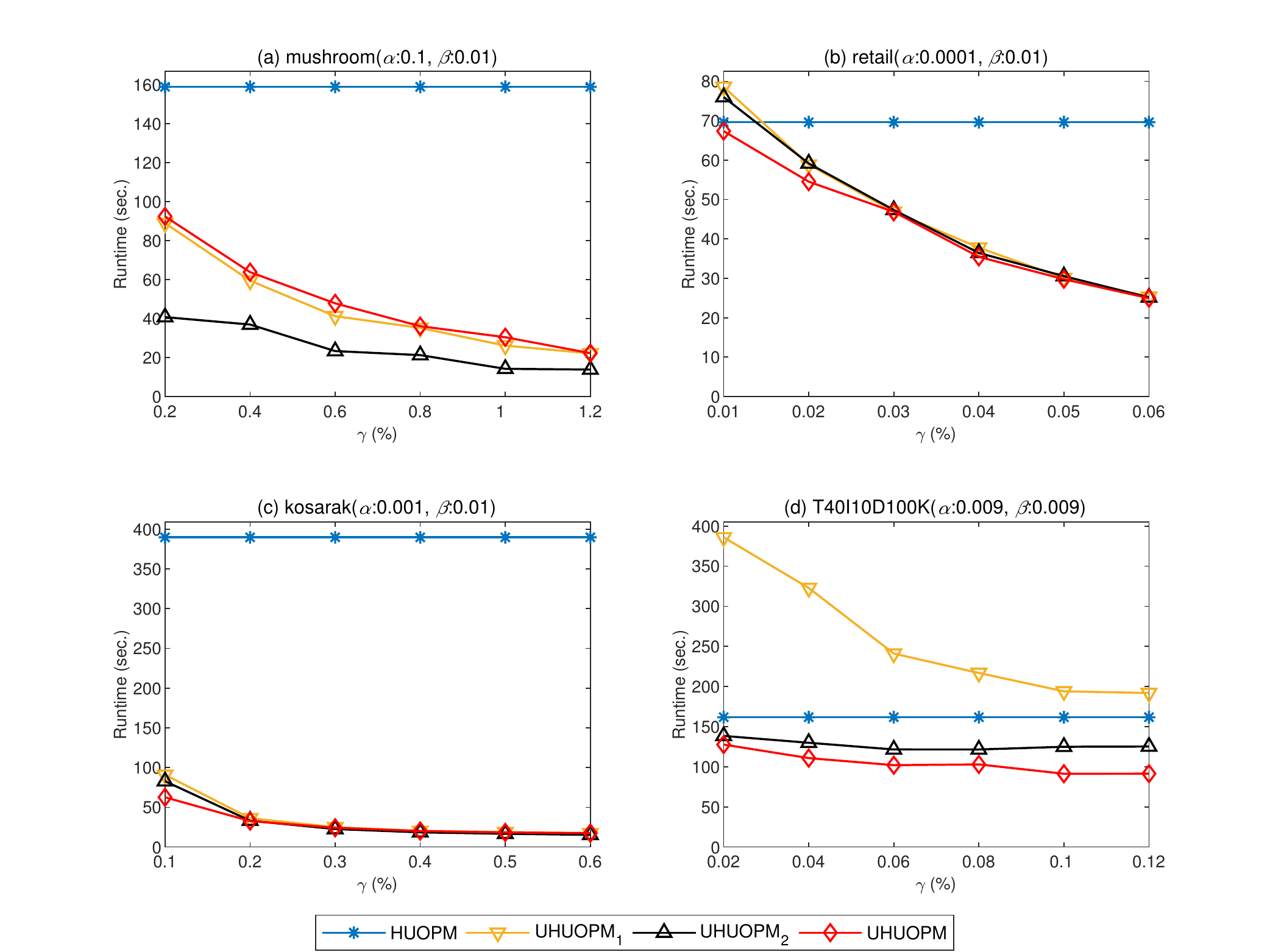}
	\captionsetup{justification=centering}
	\caption{Runtime under a changed $\gamma$ with a fixed $\alpha$ and $\beta$}
	\label{fig:runtimeMP}	
\end{figure}
% % % % % % % % % % % % % % %

The trends of the runtime under the conditions that the support, utility occupancy, and probability separately changed while the other two parameters were fixed are respectively shown in Fig. \ref{fig:runtimeMS}, Fig. \ref{fig:runtimeMU}, and Fig. \ref{fig:runtimeMP}. Since the HUOPM algorithm is based on an precise dataset, it does not contain probability values.

The UHUOPM algorithm was generally superior to the other three algorithms in runtime except for on the mushroom dataset, as shown in Fig. \ref{fig:runtimeMS}, Fig. \ref{fig:runtimeMU}, and Fig. \ref{fig:runtimeMP}. This implies that several of the strategies proposed in this paper worked well. For example, as shown in \ref{fig:runtimeMS} (d), $\beta$ was set to 0.01, $\gamma$ was 0.001, and $\alpha$ changed from 0.9$\%$ to 1$\%$ in increments of 0.02$\%$. It can be seen from the figure that the runtime of the HUOPM algorithm was the longest, and the runtime of the UHUOPM algorithm was the shortest. The runtime of the other two versions of the proposed algorithm were between HUOPM and UHUOPM. This is due to the lack of probability constraints in HUOPM. The number of traversed nodes was much more than the other algorithms, and thus the time consumption was significantly large. Compared to the UHUOPM algorithm, UHUOPM$_1$ and UHUOPM$_2$ were not good enough because they used part of the proposed strategies. They traversed more nodes, and thus their runtimes were slightly more than the UHUOPM algorithm. As shown in Fig. \ref{fig:runtimeMS}(a), $\beta$ was set to 0.1, $\gamma$ was 0.01, and $\alpha$ changed from 9$\%$ to 14$\%$ in increments of 1$\%$. This figure shows that the runtime of the UHUOPM$_2 $ algorithm was the shortest. This is because the mushroom dataset is very dense, and Strategy \ref{stra_1} and Strategy \ref{stra_3} play an obvious role while Strategy \ref{stra_2} and Strategy  \ref{stra_4} had little effect. A similar situation occurred when the utility occupancy was fixed or the probability was fixed, as shown in the other datasets. Besides, Fig. \ref{fig:runtimeMP} depicts the condition in the runtime when the other parameters were fixed and the probability varied. No matter how the minimum probability threshold varied, the runtime of the HUOPM algorithm was always stable. This is because the test datasets involved in the HUOPM algorithm were precise instead of uncertain. In other words, all the occurrence probabilities of the itemsets processed by HUOPM were 1.0; thus, its image curve was reasonably a straight line.

% % % % % % % % % % % % % % %
\begin{figure}[!hbt]
	\centering %所给出的四个数字 分别代表了从左、下、右、上被截去的值
 	\includegraphics[trim=50 0 30 0,clip,scale=0.5]{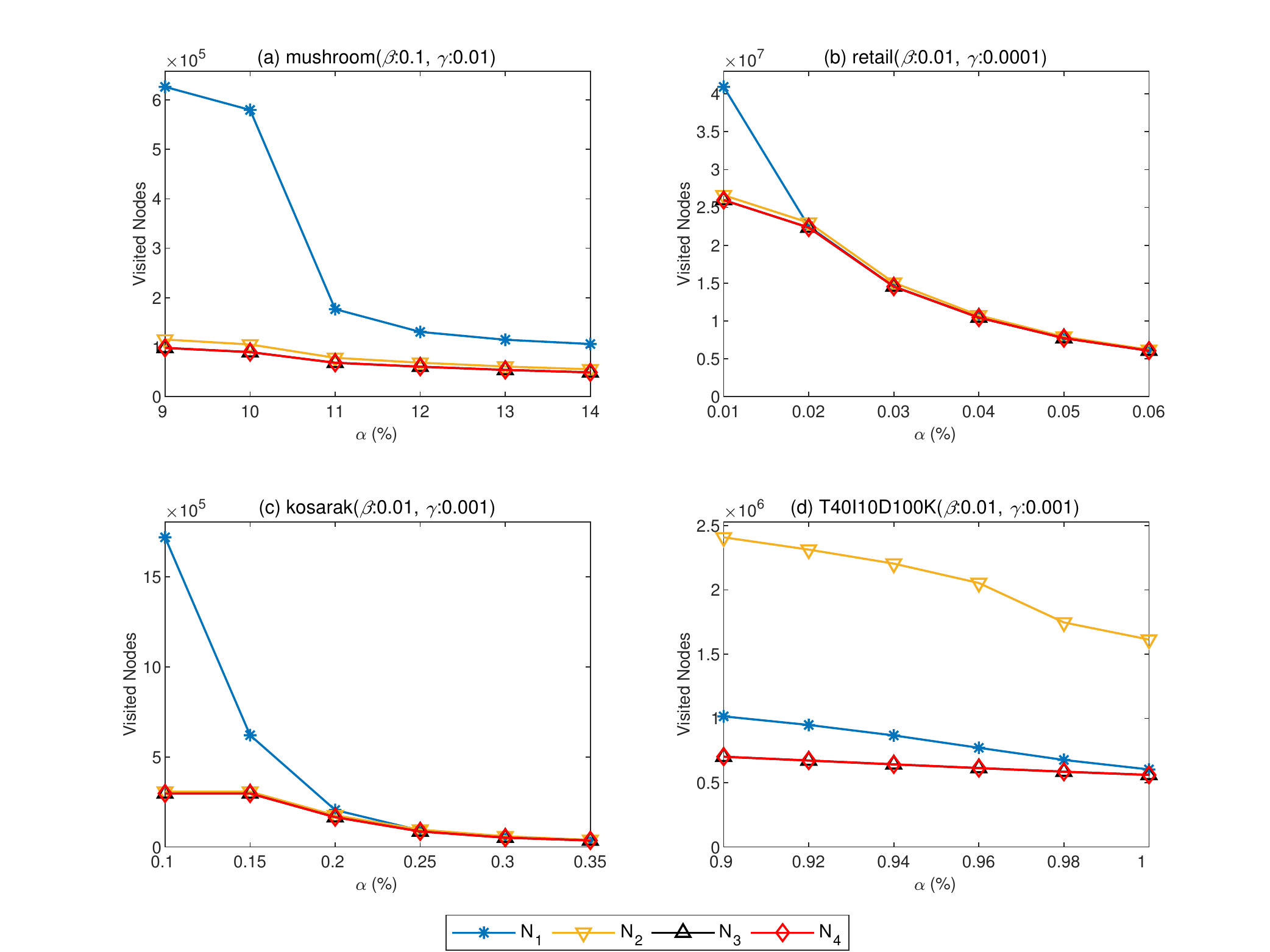}
	\captionsetup{justification=centering}
	\caption{Memory under a changed $\alpha$ with a fixed $\beta$ and $\gamma$}
	\label{fig:memoryMS}	
\end{figure}
% % % % % % % % % % % % % % %

% % % % % % % % % % % % % % %
\begin{figure}[!hbt]
	\centering %所给出的四个数字 分别代表了从左、下、右、上被截去的值
 	\includegraphics[trim=50 0 30 0,clip,scale=0.5]{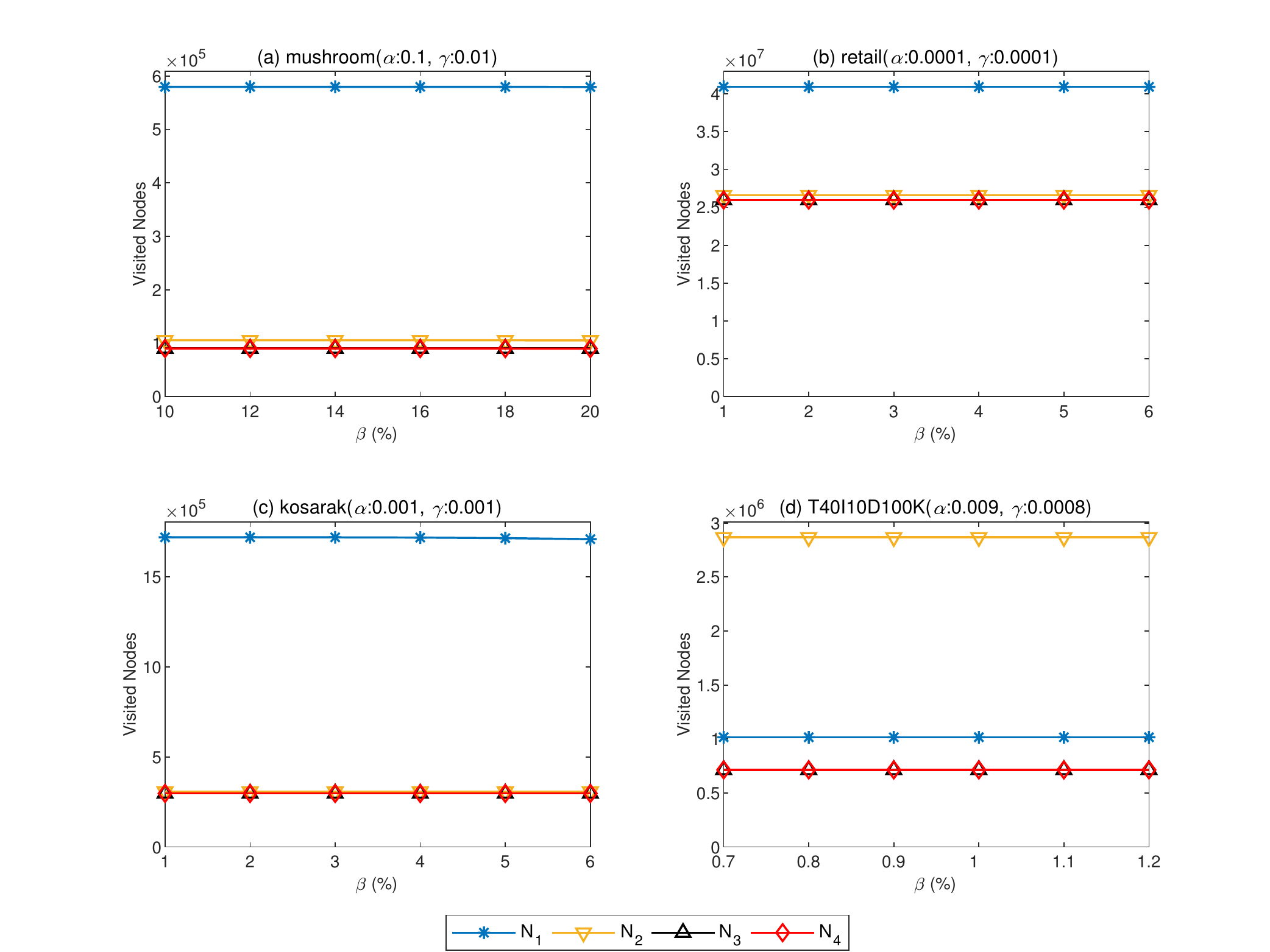}
	\captionsetup{justification=centering}
	\caption{Memory under a changed $\beta$ with a fixed $\alpha$ and $\gamma$}
	\label{fig:memoryMU}	
\end{figure}
% % % % % % % % % % % % % % %

% % % % % % % % % % % % % % %
\begin{figure}[!hbt]
	\centering %所给出的四个数字 分别代表了从左、下、右、上被截去的值
 	\includegraphics[trim=50 0 30 0,clip,scale=0.5]{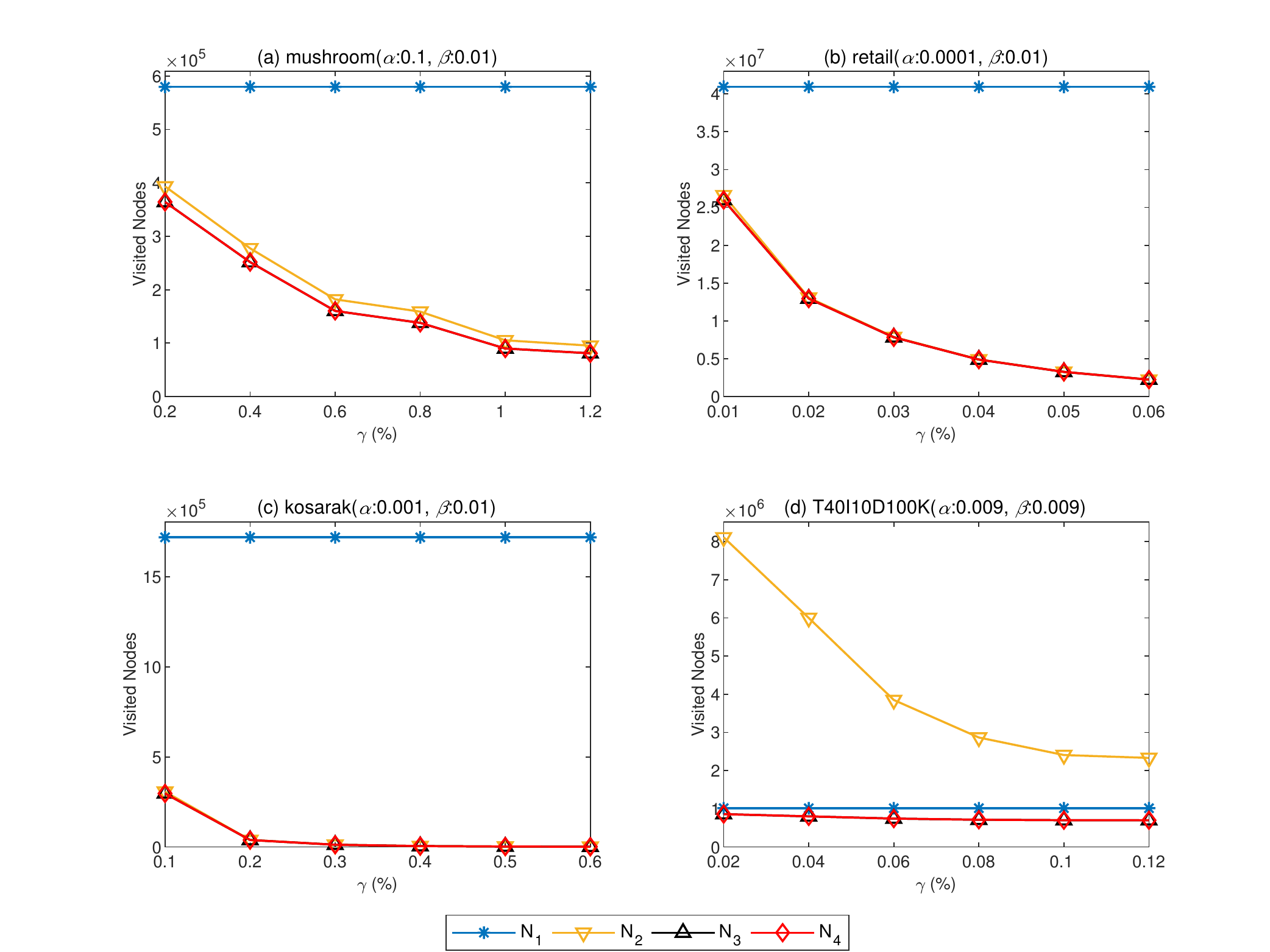}
	\captionsetup{justification=centering}
	\caption{Memory under a changed $\gamma$ with a fixed $\alpha$ and $\beta$}
	\label{fig:memoryMP}	
\end{figure}
% % % % % % % % % % % % % % %

% \subsection{Visited Nodes Analysis}
% \subsection{Memory usage analysis}
\subsection{Visited node analysis}

Because the interesting patterns needed to be saved into the memory during the algorithm execution, although the information has developed rapidly in the big data era, many large-capacity storage media could be found, but the demand for memory consumption was still large. Hence, for data mining tasks, it is a common demand to reduce memory usage. This subsection is mainly used to compare the number of nodes visited by several algorithms. When each node in the search space is accessed, the corresponding PUO-list and PFO-table should be constructed and they need to consume a certain amount of memory space. Thus, the detailed memory consumption of these algorithms can be indirectly reflected by measuring the number of visited nodes. For the convenience of observation, let the number of nodes visited by the four compared algorithms be $N_1$, $N_2$, $N_3$, and $N_4$. The experimental comparisons are shown in Fig. \ref{fig:memoryMS}, Fig. \ref{fig:memoryMU}, and Fig. \ref{fig:memoryMP}, respectively.

It is obvious that whether the support, utility occupancy, or probability varied, the UHUOPM algorithm required less memory consumption compared to the UHUOPM$_1 $ and UHUOPM$_2 $ algorithms, both of which adopted partial pruning strategies and had less-visited nodes than that of the state-of-the-art HUOPM algorithm for the selected four datasets under different characteristics. For example, in Fig. \ref{fig:memoryMP} (d), $\alpha$ was set to 0.9$\%$, $\beta$ was also set to 0.9$\%$, and $\gamma$ increased from 0.02$\%$ to 0.12$\%$ in increments of 0.02$\%$. With the gradual increase of $\gamma$, all four polylines show a downward trend, which means that as the value of $\gamma$ increased, the constraint of the probability conditions on the traversal nodes also increased accordingly, which naturally reduced the number of nodes that met the conditions of derived interesting patterns. However, without the constraint of probability, the number of visited nodes in the HUOPM algorithm was significantly more than that of the UHUOPM algorithm, and was sometimes even dozens of times.

\subsection{Patterns analysis}

The PHUOPs mined by the proposed algorithm in the uncertain datasets are further evaluated in this section. Since no existing methods have been proposed in the literature for discovering the potential high utility-occupancy patterns from uncertain datasets, the state-of-the-art HUOPM was chosen for comparison with the algorithms mentioned in this paper. Although UHUOPM$_1 $ and UHUOPM$_2 $ only have some of the pruning strategies, they had the same restrictions on the patterns. Therefore, they could successfully discover the same number of target patterns as the UHUOPM algorithm. Based on this, we only compared the number of patterns found by the HUOPM algorithm and the UHUOPM algorithm on four different datasets and recorded them as $N_1$ and $N_2$, respectively.

% % % % % % % % % % % % % % %
\begin{figure}[!hbt] 
	\centering %所给出的四个数字 分别代表了从左、下、右、上被截去的值
 	\includegraphics[trim=60 0 30 0,clip,scale=0.52]{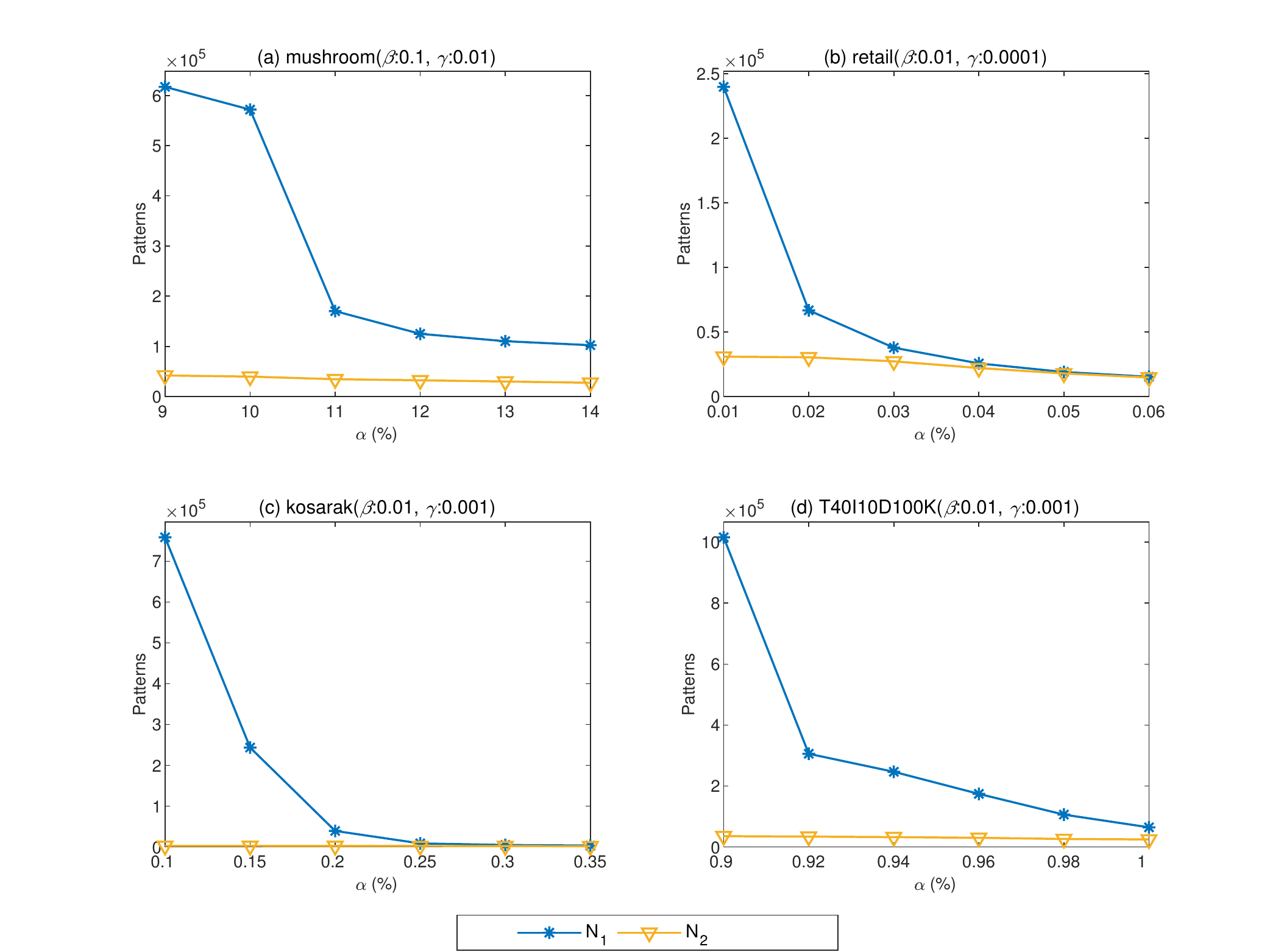}
	\captionsetup{justification=centering}
	\caption{Patterns under a changed $\alpha$ with a fixed $\beta$ and $\gamma$}
	\label{fig:patternMS}	
\end{figure}
% % % % % % % % % % % % % % %

% % % % % % % % % % % % % % %
\begin{figure}[!hbt]
	\centering %所给出的四个数字 分别代表了从左、下、右、上被截去的值
 	\includegraphics[trim=60 0 30 0,clip,scale=0.52]{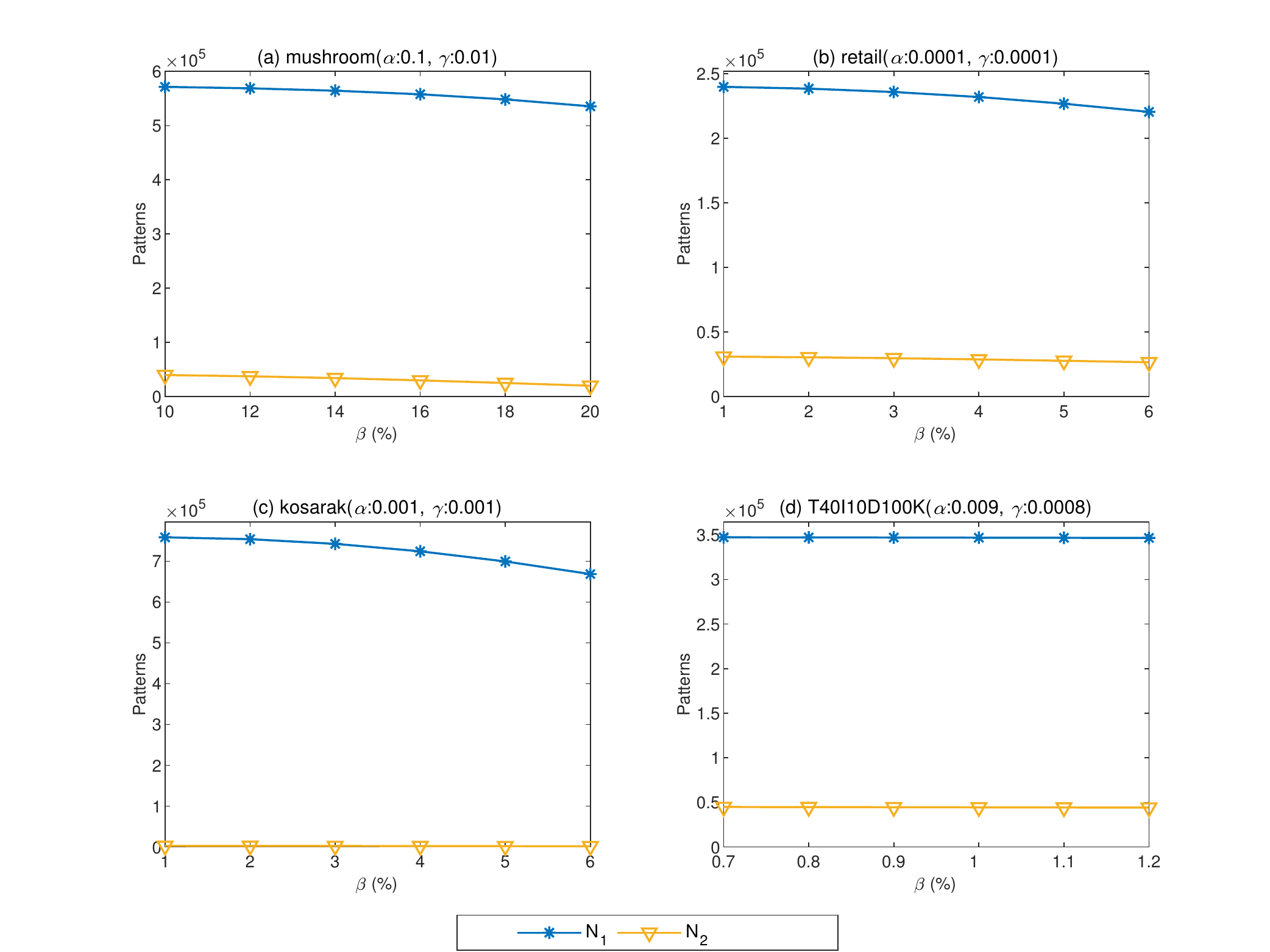}
	\captionsetup{justification=centering}
	\caption{Patterns under a changed $\beta$ with a fixed $\alpha$ and $\gamma$}
	\label{fig:patternMU}	
\end{figure}
% % % % % % % % % % % % % % % 	

% % % % % % % % % % % % % % %
\begin{figure}[!hbt]
	\centering %所给出的四个数字 分别代表了从左、下、右、上被截去的值
 	\includegraphics[trim=60 0 30 0,clip,scale=0.52]{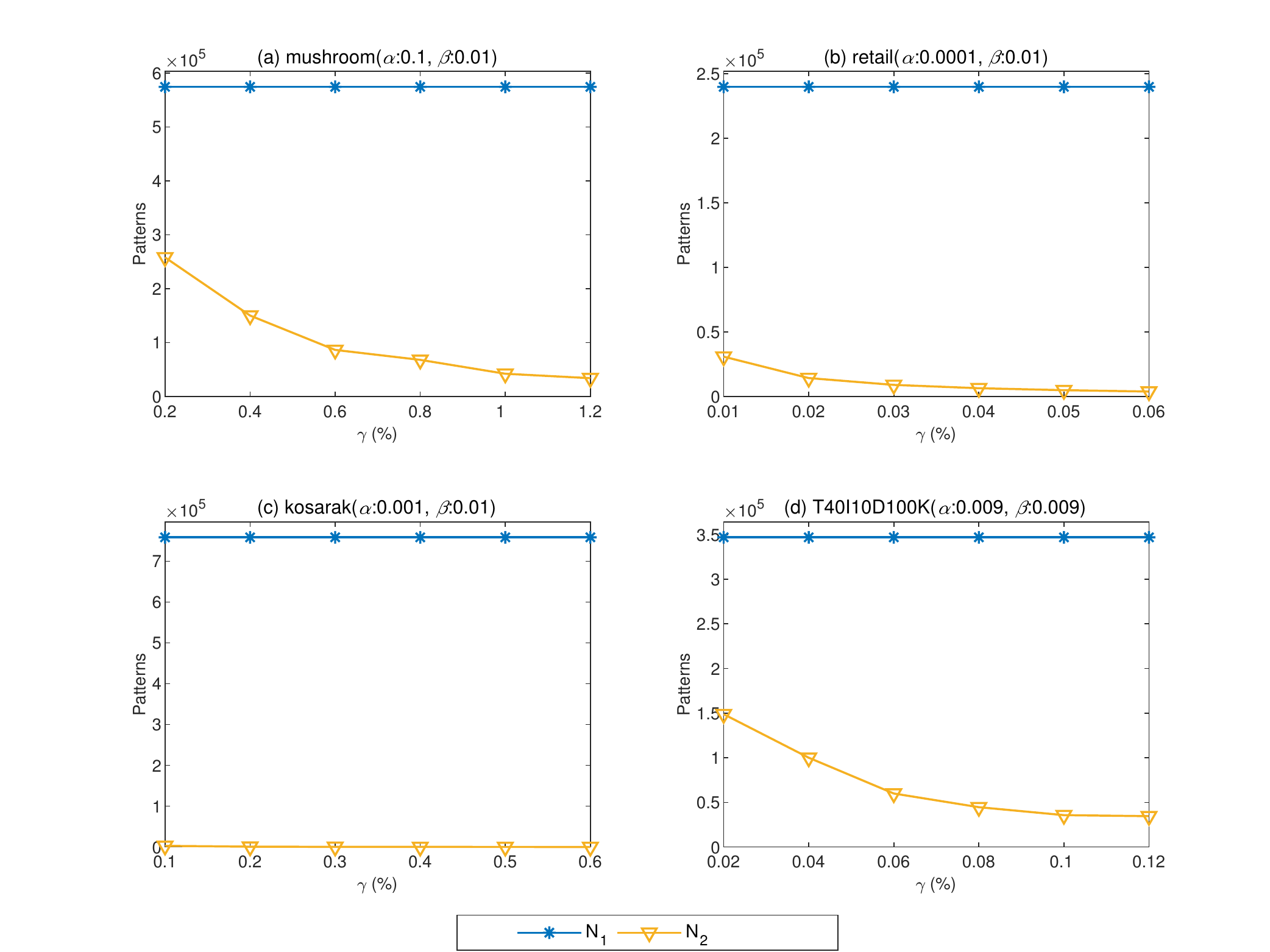}
	\captionsetup{justification=centering}
	\caption{Patterns under a changed $\gamma$ with a fixed $\alpha$ and $\beta$}
	\label{fig:patternMP}	
\end{figure}
% % % % % % % % % % % % % % % 

A comparison of the number of valid patterns as the $\alpha$, $\beta$, and $\gamma$ changed is shown in Fig. \ref{fig:patternMS}, Fig. \ref{fig:patternMU}, and Fig. \ref{fig:patternMP}. These figures show that the number of potential high-utility occupancy patterns found in an uncertain dataset should be less than the number found in a precise dataset, and sometimes it could be up to ten times less.

For example, Fig. \ref{fig:patternMS} (a) shows that as the minimum threshold continued to increase, the number of patterns found by both algorithms constantly decreased. Furthermore, few PHUOPs were always discovered on the given datasets. Moreover, compared to the HUOPM algorithm, the line chart of the UHUOPM algorithm was more stable. This is reasonable because it not only considers the utility occupancy and support restrictions in mining PHUOPs but also the role of probability.

\section{Conclusion and Future Work} %  and Future Works
\label{sec:conclusion}

%  and can be referred to as UHUOPM

So far, many algorithms have been proposed to solve the problem of extracting high utility patterns in precise quantitative databases or mining frequent patterns in uncertain databases. However, there has still been no algorithm proposed to discover potential high utility-occupancy patterns in uncertain databases. To solve this problem, a novel algorithm, namely UHUOPM, was proposed in this paper. The proposed algorithm adopts a novel mining framework that uses two list-based structures to reduce database traversal. Several efficient pruning strategies were utilized to improve the efficiency of searching and reduce the running time. Follow-up experiments were conducted to analyze the performance of the compared algorithms in terms of runtime, visited nodes w.r.t. memory consumption, and found patterns. Since this is the first work to find PHUOPs in an uncertain database, there is still much room for future research in terms of different constraint-based patterns or other types of processed data.

\section*{Acknowledgment}
This work is supported in part by the National Natural Science Foundation of China under Grant 61300167 and Grant 61976120, the Natural Science Foundation of Jiangsu Province under Grant BK20151274 and Grant BK20191445, and the Six Talent Peaks Project of Jiangsu Province under Grant XYDXXJS-048.

%\section*{References}
\bibliography{main}

\end{document}